\documentclass[11pt]{article}

\usepackage{authblk}

\usepackage{amsmath,amssymb,amsthm}
\usepackage{booktabs}
\usepackage{multirow}
\usepackage{makecell}
\usepackage{enumerate}
\usepackage{graphicx}
\graphicspath{{figs/}}
\usepackage{url}
\usepackage{color,subcaption}
\usepackage{geometry}
\usepackage{algorithm}
\usepackage{algpseudocode}
\usepackage{caption}
\usepackage[counterclockwise]{rotating}
\usepackage[T1]{fontenc}
\usepackage{setspace}

\usepackage{thmtools}
\usepackage{thm-restate}

\usepackage[bookmarksopen=false]{hyperref}

\usepackage[flushleft]{threeparttable}

\usepackage[round]{natbib}

\usepackage{tikz}
\hypersetup{pdfborder = {0 0 0},colorlinks=true,linkcolor=blue,citecolor=blue,urlcolor=black}
\usepackage{plain}
\usepackage{amsfonts}
\usepackage{color}
\usepackage{algorithm}
\usetikzlibrary{positioning} 
\usepackage{xcolor} 
\usepackage{lineno} 

\newtheorem{theorem}{Theorem}

\newtheorem{lemma}{Lemma}

\theoremstyle{definition}

\newcommand{\bZ}{{\mathbf{Z}}}

\newcommand{\bU}{{\mathbf{U}}}

\newcommand{\bX}{{\mathbf{X}}}

\newcommand{\bx}{{\mathbf{x}}}
\newcommand{\bu}{{\mathbf{u}}}

\newcommand{\btheta}{{\boldsymbol{\theta}}}

\newcommand{\beps}{{\boldsymbol{\epsilon}}}

\newcommand{\bet}{{\boldsymbol{\eta}}}

\newcommand{\bSigma}{{\boldsymbol{\Sigma}}}

\newcommand{\tabincell}[2]{\begin{tabular}{@{}#1@{}}#2\end{tabular}}

\title{Covariate balancing with measurement error}

\author{Xialing Wen\thanks{Email: \href{mailto: wenxl3@mail2.sysu.edu.cn}{wenxl3@mail.sysu.edu.cn}}\ \ \ and \
Ying Yan\thanks{Corresponding Author. Email: \href{mailto: yanying7@mail.sysu.edu.cn}{yanying7@mail.sysu.edu.cn}}}

\affil{School of Mathematics, Sun Yat-sen University, No. 135 Xingang Xi Road, Guangzhou 510275, China}

\date{}

\begin{document}


\onehalfspacing

\maketitle
\thispagestyle{empty}

\begin{abstract}
\noindent
In recent years, there is a growing body of causal inference literature focusing on covariate balancing methods. These methods eliminate observed  confounding by equalizing covariate moments between the treated and control groups. The validity of covariate balancing  relies on an implicit assumption that all covariates are accurately measured, which is frequently violated  in observational studies. Nevertheless, the impact of measurement error on covariate balancing is unclear, and there is no existing work on   balancing mismeasured covariates adequately. In this article, we show that naively ignoring measurement error  reversely increases the magnitude of covariate imbalance and induces bias to treatment effect estimation. We  then propose a class of measurement error correction strategies for the existing covariate balancing methods. Theoretically, we show that these strategies  successfully recover   balance for all covariates, and eliminate bias of   treatment effect estimation. We assess the proposed correction methods in simulation studies and  real data analysis.

\end{abstract}

\newpage

\pagenumbering{arabic}
\section{Introduction \label{sec1}}
A   primary goal of causal inference in observational studies is to  infer  treatment effect on an outcome after making  adjustment for relevant covariates.
Propensity score weighting  (PSW) \citep{rosenbaum1983central,Robins2020WhatIf} is a popular method to adjust for observed confounders, where the propensity score is usually estimated through either a parametric or nonparametric model \citep{hirano2003efficient}. Nevertheless, PSW is susceptible to model misspecification and extreme weights \citep{kang2007demystifying}, and covariate imbalance may not be eliminated  after reweighting.
	
In recent years, various covariate balancing methods have  been developed as an alternative framework to PSW. These methods  achieve finite-sample covariate balance  by directly equalizing  the weighted covariate moments across different treatment groups. For example,
	\cite{hainmueller2012entropy} proposed entropy balancing (EB), which formulates a constrained optimization problem using entropy as the loss function subject to a set of balancing constraints.
	\cite{zhao2016entropy} showed that EB possesses the double robustness property.
	Recently, EB has been  adapted to address the transportation and generalizability issues in data integration \citep[e.g.,][]{dong2020integrative}.
	There is a rich literature  related to the EB framework which explored various loss functions, inequality constraints, and other issues \citep{chan2016globally, Zhao2019, wang2020minimal, tan2020regularized, Josey2021,DaiYan2022}. In contrast to EB, \cite{imai2014covariate} proposed the covariate balancing propensity score (CBPS) framework, where a parametric propensity score model is fitted subject to some balancing constraints. The CBPS framework has undergone further development, including extension to handle   longitudinal data and continuous treatment \citep{ImaiRatkovic2015, FongHazlettImai2018, ZhouWodtke2020, Kallus2021, fan2016improving}.
Nevertheless,  the validity  of the aforementioned covariate balancing methods relies on an implicit assumption that all adjusted covariates are precisely measured.
	
Measurement error and misclassification are common challenges in data collection and analysis \citep{Carroll, Yibook}. In observational studies,
	\cite{McCaffrey} and \cite{Lockwood} showed that naively ignoring measurement error in  PSW and matching  leads to nonignorable bias of causal effect estimation and inference even when the sample size approaches infinity. This type of bias is often referred to as measurement bias or information bias in the causal inference literature  \citep{Robins2020WhatIf}.  The causal pathway from the observed mismeasured covariates to the treatment   and the outcome  is mediated indirectly through the underlying true covariates, and thus adjusting for the mismeasured covariates, in comparison of omitting these covariates, may reduce bias in estimating treatment effects \citep{maathuis2018handbook}. Nevertheless, the bias can be still substantial, because the imperfect measurements alone do not formulate a sufficient adjustment set \citep{ogburn2013bias}.  Measurement correction strategies have been developed for PSW to completely eliminate  measurement bias under the classical additive measurement error model  \citep{McCaffrey,YanRen}. To the best of our knowledge, there is no existing  work  that reveals the impact of measurement error on covariate balance and makes proper measurement error correction for the  balancing methods.
	
In this article, we conduct a systematic study on  measurement error problems for covariate balancing. The contribution is two-fold. First, we show that naively ignoring measurement error and applying the  covariate balancing methods  reversely increases the magnitude of covariate imbalance and induces measurement bias in  causal effect estimation.  In particular, we derive bias formulas and tight bounds. Second, we proceed to address the challenge of restoring covariate balance  and establishing credible causal inference in the presence of measurement error. Specifically, we  propose a class of measurement error correction strategies for the existing covariate balancing methods, either imposing a parametric
assumption for the measurement error distribution or relaxing the distributional assumption when replicated measurements are available.
We prove that these strategies successfully recover covariate
balance and remove bias in causal effect estimation. Asymptotic normality and double robustness properties are established.

The paper is organized as follows. In Section \ref{sec2}, we introduce  notations and setup. In Section \ref{sec3}, we derive bias formulas and bounds to quantify measurement bias in covariate balance and treatment effect estimation incurred by naive  covariate balancing. In Section \ref{sec4}, we present measurement error correction strategies. In Section \ref{sec5}, we conduct simulation studies and real data analysis. The Supplementary Material includes  regularity conditions and proofs.

\section{Notation and Preliminary \label{sec2}}

Consider a  random sample  of $n$ subjects from a target population. Each subject has an independent copy of the triplet $(T, \bZ, Y)$, where $\bZ$ is a $p$-dimensional vector of baseline covariates,  $T$ is a binary treatment indicator with $T=1$ for the treated and $T=0$ for the control, and  $Y$ is an outcome variable. We write the  data set to be $\mathcal{O} = \{(T_{i}, \bZ_{i}, Y_{i}): i=1, \ldots, n\}$, where the treated and control groups consists of   $n_{1}=\sum_{i=1}^n T_i$ and $n_{0}=n-n_1$ subjects, respectively.
Let $Z_j$ be the $j$-th component of $\bZ$, and $Z_{ij}$ be the corresponding component for the $i$th subject, $i=1,\cdots,n$, $j = 1, \cdots, p$.
Following the potential outcome framework \citep[e.g.,][]{imbens2015causal}, let $Y(0)$ and $Y(1)$ be the potential outcomes under treatments $T=0$ and $T=1$, respectively. The observed outcome is $Y = (1-T) Y(0) + T Y(1)$.

Our analysis relies on two critical assumptions: (i) (strong ignorability) $\{Y(0), Y(1)\} \perp T \mid \bZ$, and (ii) (positivity) $0 < \text{Pr}(T=1 \mid \bZ) < 1$ over the support of $\bZ$. Throughout the paper, the causal estimand of interest is the average treatment effect on the treated (ATT), denoted by $\tau = E\{Y(1)-Y(0)\mid T=1\}$. The conclusions of the paper can be readily  extended to other estimands, including the average treatment effect and the average treatment effect on the control.

We consider a general  scheme where  a set of weights $\{\hat{w}_i; \ T_i=0\}$ for the control group are constructed using the data subset $\mathcal{O}_1=\left\{(T_{i}, \bZ_{i}): i=1, \cdots, n\right\}$ where the outcome information is excluded. In general, a  weighting  estimator for  ATT is given by
\begin{equation}
\hat{\tau}=\bar{Y}^{trt} - \sum\limits_{i: T_i=0} \hat{w}_{i} Y_{i},\label{genericATT}
\end{equation}
where $\bar{Y}^{trt}=\sum_{i: T_i=1}  Y_{i}/n_1$ is the sample mean of the outcomes in the treated group.

\subsection{Covariate balancing Framework \label{subsec2.2}}

To ensure finite-sample covariate balance between the treated and control groups, many weighting methods directly incorporate  balancing conditions in the weight construction procedure.  We  categorize a majority of  these covariate balancing methods into two frameworks.

The first framework entails solving the following constrained convex optimization problem to obtain  balancing weights:
\begin{equation}\label{primal}
	\begin{aligned}
		&\underset{\{w_i; T_i=0\}}{\min}  \quad  \sum\limits_{i: T_i=0} f(w_{i})\\
		&	\text{s.t.}
		\left\{
		\begin{array}{lr}
			\sum\limits_{i: T_{i}=0} w_{i}{Z}_{ij}=\bar{Z}_{j}^{trt},  & j=1, \cdots, p, \\
			\sum\limits_{i: T_{i}=0} w_{i}=1, & \\
			w_{i}\geq 0, &   \text{for } i\in\{j: T_j=0\},
		\end{array} \right.
	\end{aligned}
\end{equation}
where $\bar{Z}_{j}^{trt}=\sum_{i: T_i=1} Z_{ij}/n_1$ represents the sample mean of $Z_j$ in the treated group, and  $f(\cdot)$ is a pre-specified loss function.
The balancing conditions, given by
\begin{equation}
	\sum\limits_{i: T_{i}=0} w_{i}{Z}_{ij}=\bar{Z}_{j}^{trt}, \ \ j=1, \cdots, p,
	\label{balcondition}
\end{equation}
are the moment constraints in the primal problem \eqref{primal}, which guarantee that exact covariate balance  between the two groups is achieved after reweighting.

This framework consists of EB \citep{hainmueller2012entropy,zhao2016entropy}, stable weighting \citep{zubizarreta2015stable,wang2020minimal}, empirical balancing calibration weighting \citep{chan2016globally}, and Bregman distance-based covariate balancing \citep{Josey2021}, among others. While these methods use similar constraints, they differ in  the choice of the loss function. In particular, EB employs the  entropy
loss $f(x) = x\log x$, and  the dual optimization problem  is
\begin{equation}\label{eq:dual-opt-EB}
	\begin{aligned}
	\hat{\btheta}=	\underset{{\btheta}\in \mathbb{R}^p}{\text{argmin}}  \ \log \left\{\sum\limits_{i: T_i=0} \exp \left(\btheta^{\top} \bZ_{i}\right)\right\}-\btheta^{\top}\bar{\bZ}^{trt},
	\end{aligned}
\end{equation}
where  $\btheta$ is the dual parameter.
The   EB weight  for subject $i$ in the control group is
\begin{equation}\label{EBweight}
\hat{w}_{i} = \frac{\exp({\hat{\btheta}}^{\top} \bZ_{i})}{\sum\limits_{j:T_{j}=0} \exp({\hat{\btheta}}^{\top} \bZ_{j})}.
\end{equation}
 The EB-based ATT estimator is then obtained through equation \eqref{genericATT}.

The second  framework, namely CBPS \citep{imai2014covariate}, fits a parametric propensity score model $\text{Pr}(T=1|\bZ) = \pi_{\btheta}(\bZ)$ with $p$-dimensional parameter $\btheta$  subject to a set of balancing conditions.  In specific, the CBPS method estimates $\btheta$ by solving the following over-identified estimating equations:
\begin{equation}\label{eq:cbpsbalancing}
	\begin{aligned}
&\sum_{i=1}^n \left\{\frac{T_i \pi_{{\btheta}}^\prime(\bZ_i)}{\pi_{{\btheta}}(\bZ_i)}-\frac{(1-T_i) \pi_{{\btheta}}^\prime(\bZ_i)}{1-\pi_{{\btheta}}(\bZ_i)}\right\}=\mathbf{0}_p,\\
\text{and}\ \ & \sum\limits_{i: T_i=0} w_i Z_{ij}=\bar{Z}_j^{trt} ,\ j=1,\cdots, p,
	\end{aligned}
\end{equation}
where the first set of estimating equations in \eqref{eq:cbpsbalancing} are the likelihood score equations, and the second set  consists of  balancing conditions, where $ {w}_{i}= {n_1}^{-1} \pi_{{\btheta}}(\bZ_i)\{1-\pi_{{\btheta}}(\bZ_i)\}^{-1}$ is the parameterized  weight for subject $i$ in the control group. Let $\hat{\btheta}$ be the solution of the   equations in \eqref{eq:cbpsbalancing}.
 The CBPS weight is
 \begin{equation}\label{CBPSweight}
\hat{w}_{i} = \frac{\pi_{\hat{\btheta}}(\bZ_i)}{n_1\{1-\pi_{\hat{\btheta}}(\bZ_i)\}}.
\end{equation}
We impose a logistic propensity score model $\pi_{{\btheta}}(\bZ)=\text{logit}^{-1}(\bZ^\top \btheta)$ under the CBPS framework, where $\text{logit}(\cdot)$ is the logistic function and  a constant term  is included in $\bZ_i$. Therefore, $\sum_{j:T_{j}=0} \hat{w}_{j}=1$. The CBPS-based ATT estimator is  obtained by equation \eqref{genericATT}.
  CBPS is shown to be closely related to EB \citep{zhao2016entropy}.

If  precise measurements of all covariates are available, then under the logistic propensity score model the EB estimator $ \hat{\btheta} $ is consistent of $ \btheta_0$ \citep{zhao2016entropy}. Likewise, the CBPS estimator $ \hat{\btheta} $ is consistent \citep{fan2016improving}.

\subsection{Imbalance measure \label{subsec2.im}}
We define two imbalance measures to evaluate residual imbalance after reweighting for an arbitrary weighting method. The first measure is the  absolute standardized mean difference (ASMD) \citep{imbens2015causal}:
\begin{equation}
	{ASMD}_j=\frac{\left|\bar{Z}_{j}^{trt}-\sum\limits_{i:T_i=0} \hat{w}_{i}  Z_{ij}\right|}{\hat{\sigma}_{j}^{trt}}, \ \ j=1,\cdots, p,\label{wASMD}
\end{equation}
where  $\hat{\sigma}_{j}^{trt}=\sqrt{\sum_{i:T_i=1} (Z_{ij}-\bar{Z}_{j}^{trt})^2/(n_{1}-1)}$ is the sample standard deviation of $Z_{j}$ in the treated group. The ASMD serves as a univariate measure quantifying  remaining imbalance for each covariate.

The second measure is Mahalanobis distance (MD) \citep{imai2014covariate,imbens2015causal}:
\begin{equation}\label{wMD}
	\begin{aligned}
		{MD}=&\sqrt{\left(\bar{\bZ}^{trt}-\sum\limits_{i:T_i=0}  \hat{w}_{i} \bZ_i\right)^{\top}\left(\hat{\mathbf{\Sigma}}^{trt}\right)^{-1}\left(\bar{\bZ}^{trt}-\sum\limits_{i:T_i=0}  \hat{w}_{i} \bZ_i\right)},
	\end{aligned}
\end{equation}
where $\bar{\bZ}^{trt}=\sum_{i:T_i=1}\bZ_{i} / n_1$ and $\hat{\mathbf{\Sigma}}^{trt}=\sum_{i:T_i=1} (\bZ_{i}-\bar{\bZ}^{trt})(\bZ_{i}-\bar{\bZ}^{trt})^\top /(n_{1}-1)$ are  the sample mean and sample variance-covariance matrix of $\bZ$ in the treated group, respectively. The MD is a multivariate imbalance measure of overall residual imbalance after reweighting.

In the absence of covariate measurement error,   the EB and CBPS methods  achieve \textit{finite-sample  exact balance} in the sense that for each method the following is true: ${ASMD}_j={MD}=0$, $j=1\cdots, p$.

\subsection{Measurement error \label{subsec2.3}}

In the rest of the paper, we consider the common situation that the true covariates $ \bZ $ are subject to measurement error. Without loss of generality, write $ \mathbf{Z} = (\mathbf{X}^\top, \mathbf{U}^\top)^\top $, where $ \mathbf{X} $ is a vector of $ p_1 $-dimensional error-prone covariates, and $ \mathbf{U}$ is a vector of $ p_2 $-dimensional accurately measured covariates, where $ p=p_1+p_2 $. Suppose that the true covariates $ \mathbf{X} $ are  unobservable. The imperfect measurements, denoted by $\bX^{*}$, are collected instead. Define $ \mathbf{Z}^* = (\mathbf{X}^*{}^\top, \mathbf{U}^\top)^\top $.  We assume the classical additive measurement error model \citep{Carroll}, under which the relationship between the true covariates $ \mathbf{Z} $ and the observed counterpart $ \mathbf{Z}^*$ is modeled as:
\begin{equation}
	\begin{aligned}
	\mathbf{Z}_i^* = \mathbf{Z}_i + \boldsymbol{\epsilon}_i, \ \ i=1,\cdots, n,
		\label{me}
	\end{aligned}
\end{equation}
where the mean-zero error terms $\boldsymbol{\epsilon}_i$ are mutually independent, $i=1,\cdots,n$, and they are independent of the underlying data set $\mathcal{O}$. Let ${\epsilon}_{ij}$ be the $j$th component of $\boldsymbol{\epsilon}_i$. Note that ${\epsilon}_{ij}=0$, $j=p_1+1,\cdots, p$. Let $ \boldsymbol{\epsilon}_{i\scriptscriptstyle X}=({\epsilon}_{i1},\cdots, {\epsilon}_{ip_1})^\top$. Define $\mathbf{\Sigma}_1=\text{Var}(\boldsymbol{\epsilon}_{i\scriptscriptstyle X})$ and $\mathbf{\Sigma}=\text{Var}(\boldsymbol{\epsilon}_i)$. Assume that $\bSigma_1$ is positive definite. Note that $\mathbf{\Sigma}$ is a block diagonal matrix:
$\mathbf{\Sigma} = \text{diag}(\mathbf{\Sigma}_1 ,\boldsymbol{0}_{p_2\times p_2})$.

To address measurement error effects and make appropriate corrections, a common strategy is to make a distributional assumption for the error term.
The normal error assumption is commonly imposed in the measurement error literature \citep{Carroll}, where the variance-covariance matrix $\mathbf{\Sigma}_1$ is typically taken from a pilot study, a previous study, or estimated from additional data sources. Moreover, when extra information on a validation subsample or replicated measurements  is available, the distributional assumption can be removed.  The details and implications of these strategies are  elaborated in Section \ref{sec4}.

\subsection{Naive covariate balancing  \label{subsec2.4}}

Let $ \mathcal{O}^* = \{(T_i, \mathbf{Z}_i^*, Y_i): i = 1, \ldots, n\} $ be the observed data set, and   $ \mathcal{O}_1^* = \{(T_i, \mathbf{Z}_i^*): i = 1, \ldots, n\} $  be its outcome-free data subset.  Write the parameter in the EB or CBPS framework to be $ \btheta=(\btheta_{\bx}^{\top}, {\btheta}_{\bu}^{\top})^\top $, where $ \btheta_{\bx} = (\theta_1, \cdots, \theta_{p_1})^\top $ and $ \btheta_{\bu} = (\theta_{p_1+1}, \cdots, \theta_{p})^\top $. Let  $ \btheta_0=(\btheta_{0,\bx}^{\top},\btheta_{0,\bu}^{\top})^\top $ be the true parameter value. In the following, we define  the naive EB and CBPS methods which ignore measurement error in the estimation procedure.

The naive EB estimator of $\btheta$ solves Problem \eqref{eq:dual-opt-EB} with the partly-unobserved true covariates $\bZ$ replaced by the fully-observed imperfect measurements $\bZ^*$:
\begin{equation}\label{naiveEB}
	\hat{\btheta}^*=	\underset{{\btheta}\in \mathbb{R}^p}{\text{argmin}} \ L(\boldsymbol{\theta} ; \mathcal{O}_1^*),
\end{equation}
where $L(\boldsymbol{\theta} ; \mathcal{O}_1^*)=\log \{\sum_{i: T_i=0} \exp (\btheta^{\top} \bZ_{i}^*)\}-\btheta^{\top}\bar{\bZ}^{*,trt}$.
Then, we obtain the naive EB weight:
\begin{equation*}
\hat{w}_{i}^* = \frac{\exp({\hat{\btheta}}^{\top} \bZ_{i}^*)}{\sum\limits_{j:T_{j}=0} \exp({\hat{\btheta}}^{\top} \bZ_{j}^*)}.
\end{equation*}
The  naive EB-based ATT estimator is $ \hat{\tau}^*=\bar{Y}^{trt} - \sum_{i: T_i=0} \hat{w}_{i}^* Y_{i}$. Analogously, we define the  parameter estimator $\hat{\btheta}^*$, the  weights $\hat{w}_{i}^*$, and the ATT estimator $ \hat{\tau}^*$ for the naive CBPS method.
Define $\hat{\btheta}^*=(\hat{\btheta}_{\bx}^{*\top},\hat{\btheta}_{\bu}^{*\top})^\top $. Let $ \btheta^*=({\btheta}_{\bx}^{*\top},{\btheta}_{\bu}^{*\top})^\top $ be the asymptotic limit of $\hat{\btheta}^*$. In general,  ${\btheta}^*\neq \btheta_0$ in the presence of measurement error.

In the next section,   we show that finite-sample  covariate balance  no longer holds for all covariates by naive EB or CBPS, and  bias is introduced to the estimation of the ATT estimand. Moreover, we derive a bias formula that highlights the discrepancy between ${\btheta}^*$ and $ \btheta_0$. This result is crucial for understanding the magnitude of measurement error effect on  treatment effect estimation.

\section{Impact of Measurement Error \label{sec3}}

In this section, we analyze the impact of naively ignoring measurement error  when  the EB or CBPS  method is employed. In Section \ref{subsec3.1}, we investigate the measurement error effect on covariate balance, and discover that ignoring measurement error induces imbalance of the unobservable true covariates  $ \mathbf{X} $. We quantify the magnitude of covariate imbalance using asymptotic analysis. In Section \ref{subsec3.2}, we show that  covariate imbalance induced by measurement error leads to biased treatment effect estimation.

\subsection{Covariate imbalance \label{subsec3.1}}

In this subsection, we  study the consequence of disregarding measurement error on covariate balance. The  error-prone covariates $ \bZ^* $ are exactly balanced for the naive EB and CBPS methods  in Section \ref{subsec2.4}, because  the balancing conditions for   $ \bZ^* $, that is,
$\sum_{i: T_i=0} w_i Z_{ij}^*=\bar{Z}_j^{*,trt},  \ \ j=1, \cdots, p$,
are directly imposed in the estimation procedure for each method. However, it is critical to note that naive covariate balancing  does not equalize sample moments for the underlying true covariates $ \bZ $ across the treated and  control groups.

To evaluate the performance of  naive EB and CBPS in balancing $ \bZ $, we modify the ASMD and MD imbalance measures   in \eqref{wASMD} and  \eqref{wMD}. Specifically, let
\begin{equation}\label{nEBASMD}
{ASMD}_j^*=\frac{\left|\bar{Z}_{j}^{trt}-\sum\limits_{i:T_i=0} \hat{w}_{i}^*  Z_{ij}\right|}{\hat{\sigma}_{j}^{trt}},  \ \ j=1, \cdots, p,
\end{equation}
where $\hat{w}_{i}^*$ is the naive EB or CBPS  weight defined in Section \ref{subsec2.4}. Similarly, define ${MD}^*$ to be ${MD}$ with $\hat{w}_{i}$ replaced by $\hat{w}_{i}^*$.

It is straightforward  that ${ASMD}_j^*\equiv 0, j=p_1+1, \cdots, p$, suggesting that the precisely measured covariates $ \bU=(Z_{p_1+1},\cdots, Z_{p})^\top $ are exactly balanced for the naive EB and CBPS methods. Nevertheless, the following theorem emphasizes that the unobserved true covariates $ \bX=(Z_{1},\cdots, Z_{p_1})^\top $ remain imbalanced after  reweighting, and the magnitude of covariate imbalance hinges on measurement error.

Let $ M(\btheta) = E\{\exp(\btheta^\top \beps)\} $ be the moment generating function (MGF) of the error term $ \beps $. Note that $ M(\btheta)= M_x(\btheta_{\bx}) $, where $ M_x(\btheta_{\bx})=E\{\exp(\btheta_{\bx}^\top \beps_{\bx})\} $ is the MGF of the error $ \beps_{\bx} $. The gradient of $ M(\btheta) $ is $ \nabla M(\btheta)=(\frac{\partial M_x(\btheta_{\bx})}{\partial\theta_1}, \cdots, \frac{\partial M_x(\btheta_{\bx})}{\partial\theta_{p_1}}, 0, \cdots, 0)^\top $. Let $ [\nabla M(\btheta)]_j $ represent the $ j $th element of $ \nabla M(\btheta) $. Let ${\sigma}_{j}^{trt}$ be the population standard deviation of $Z_{j}$ in the treated group, and ${\mathbf{\Sigma}}^{trt}$ be  the  population variance-covariance matrix of $\bZ$ in the treated group.

\begin{theorem}\label{thm1}
	Under the additive measurement error model \eqref{me}, it holds that as $n \rightarrow \infty$,
	$$
	\begin{aligned}
		 A S M D_j^* &\stackrel{p}{\longrightarrow}\left(\sigma_j^{\text {trt }}\right)^{-1} M^{-1}\left(\boldsymbol{\theta}^*\right)\left|\left[\nabla M\left(\boldsymbol{\theta}^*\right)\right]_j\right|,\ \ j=1, \cdots, p_1, \\
		 A S M D_j^* &= 0, \ \ j=p_1+1, \cdots, p, \\
	\text{and}	 \quad M D^* &\stackrel{p}{\longrightarrow}\left\|\left(\boldsymbol{\Sigma}^{\text {trt }}\right)^{-1 / 2} M^{-1}\left(\boldsymbol{\theta}^*\right) \nabla M\left(\boldsymbol{\theta}^*\right)\right\|_2.
	\end{aligned}
	$$
\end{theorem}

Theorem \ref{thm1} quantifies  residual covariate imbalance after reweighting using either the naive EB or CBPS method when the sample size approaches infinity. It shows that although the naive methods enforce \textit{finite-sample balance} for the precisely measured covariates $ \bU $, they do not balance the unobserved true covariates $ \bX $. The degree of covariate imbalance of $ \bX $ depends on the measurement error distribution through the MGF $M(\cdot)$. In addition, it depends on the asymptotic limit of naive parameter estimation $\boldsymbol{\theta}^*$. Under the normal error assumption,  the limit of $A S M D_j^*$ is $(\sigma_j^{\text {trt }})^{-1}|[{\bSigma_1\btheta_{\bx}^*}]_j| $, $ j=1,\cdots, p_1 $. It suggests that except the special case where $ \btheta_{\bx}^* $ equals to zero, the magnitude of  covariate imbalance increases as  measurement error becomes more prominent.

The underlying true covariates $\bZ$ play the key role of confounders in estimating  the  effect of the treatment $T$ on the outcome $Y$. Because the naive methods do not well balance $\bZ$ as delineated in  Theorem \ref{thm1},  treatment effect estimation  is expected to be biased. We study this problem focusing on naive EB in the following subsection. The discussion of naive CBPS is relegated to the Supplementary Material.

\subsection{Biased parameter and treatment effect estimation \label{subsec3.2}}

We  examine the bias of the naive EB estimator $\hat{\btheta}^*$ in equation \eqref{naiveEB}.
Consider a modified EB  estimator $ \tilde{\btheta}$:
\begin{equation}
	\tilde{\btheta} = \underset{{\btheta}\in \mathbb{R}^p}{\text{argmin}}  \left\{L(\btheta; \mathcal{O}_1) + \log M(\btheta) \right\},
	\label{EB2}
\end{equation}
where $L(\boldsymbol{\theta} ; \mathcal{O}_1)=\log \{\sum_{i: T_i=0} \exp (\btheta^{\top} \bZ_{i})\}-\btheta^{\top}\bar{\bZ}^{trt}$. This estimator is unavailable in practise because it  involves the unobservable true covariates.  The only difference between $\tilde{\btheta}$ and  the EB estimator $\hat{\btheta}={\text{argmin}}_{{\btheta}} L(\boldsymbol{\theta} ; \mathcal{O}_1)$ in equation \eqref{eq:dual-opt-EB} is that   $\log M(\btheta)$ is included in the objective function \eqref{EB2}. The EB estimator $\hat{\btheta}$ is consistent of ${\btheta}_0$ under the logistic propensity score model or the linear outcome model \citep{zhao2016entropy}. It follows immediately that in general the modified EB estimator $\tilde{\btheta}$ is inconsistent in the presence of covariate measurement error.

The following lemma states that  the modified EB estimator $\tilde{\btheta}$ and the naive EB estimator $\hat{\boldsymbol{\theta}}^*$ possess
  the same level of asymptotic bias. This lemma holds without assuming the  logistic propensity score model or the linear outcome model.

\begin{lemma}\label{lem1}
	Under the additive measurement error model  \eqref{me}, it holds that
	$$
	\hat{\boldsymbol{\theta}}^*-\tilde{\boldsymbol{\theta}} \stackrel{p}{\longrightarrow} \boldsymbol{0}_p, \text { as } n \rightarrow \infty .
	$$
\end{lemma}

Lemma \ref{lem1} implies that $\tilde{\boldsymbol{\theta}} \stackrel{p}{\longrightarrow} \btheta^* \text { as } n \rightarrow \infty$, where $\btheta^*$ is the asymptotic limit of the naive EB estimator $\hat{\boldsymbol{\theta}}^*$ defined  in Section \ref{subsec2.4}. Therefore,
the asymptotic bias  of the modified  EB estimator $\tilde{\boldsymbol{\theta}} $  is the same as that of $\hat{\boldsymbol{\theta}}^*$, which is equal to $\btheta^{*}-\btheta_0$.

For normally distributed errors, $\log M(\btheta) = \frac{1}{2}\btheta^\top\mathbf{\Sigma}\btheta = \frac{1}{2}\btheta_{\bx}^\top\mathbf{\Sigma}_1\btheta_{\bx}$. Therefore, the primal problem  \eqref{EB2} can be viewed as a penalized  regression problem using a  weighted $L_2$ norm as the regularizer. It follows that the modified  EB estimator $\tilde{\btheta}$, or equivalently asymptotically, the naive EB estimator $\hat{\boldsymbol{\theta}}^*$, is a shrinkage version of the EB estimator $\hat{\btheta}$.   The shrinkage effect of measurement error on naive EB estimation is in line with  the  well-known attenuation phenomenon in the measurement error literature \citep{Carroll}.

Write $\bar{\btheta}\in[\btheta_0,\btheta^{*}]$ if $\bar{\theta}_j=t_j \theta_{0,j}+(1-t_j)\theta^{*}_j$ for some $t_j\in[0,1]$, $j=1,\cdots, p$, where ${\theta}_j$ is the $j$th component of ${\btheta}$.  Write $\bar{\boldsymbol{\theta}} \in L( \boldsymbol{\theta}^* ; \boldsymbol{\theta}_0)$ if $\bar{\boldsymbol{\theta}} = (1-t) \boldsymbol{\theta}_0 + t \boldsymbol{\theta}^* $ for some $t \in[0,1]$. Let $\|\bx\|_q$ be the $L_q$ norm for a vector $\bx$, and $\|\mathbf{M}\|_q=\max_{\bx}\{\|\mathbf{M} \bx\|_q:\ \|\bx\|_q\leq 1\}$ be the induced matrix norm for a square matrix $\mathbf{M}$, where $q$ is an  integer in $[1,\infty]$. For a positive-definite $p\times p$ matrix ${\mathbf{M}}$, we express its inverse ${\mathbf{M}}^{-1}$ in the formulation of block matrices:
$$
{\mathbf{M}}^{-1} = \begin{pmatrix} { {\mathbf{M}}}_{11}^{I} &  { {\mathbf{M}}}_{12}^{I} \\  { {\mathbf{M}}}_{21}^{I} &  { {\mathbf{M}}}_{22}^{I} \end{pmatrix},$$
where ${{\mathbf{M}}}_{11}^{I}$ and $ { {\mathbf{M}}}_{22}^{I}$ are  $p_1 \times p_1$  and   $p_2 \times p_2$ sub-matrices, respectively.

We need more notations for parameter and ATT estimation. Define the Hessian matrix of $ L(\btheta; \mathcal{O}_1) $ to be $ \nabla^2 L(\btheta; \mathcal{O}_1)=\frac{\partial^2 L(\btheta; \mathcal{O}_1)}{\partial \btheta \partial \btheta^{\top}} $. It can be shown that $ \nabla^2 L(\btheta; \mathcal{O}_1) $ converges in probability to $ \mathbf{H}(\btheta) $ as $ n\rightarrow \infty$, where $\mathbf{H}(\btheta)=\nabla \mathbf{B}(\btheta)$ and $\mathbf{B}(\btheta)=\{E[\exp(\btheta^{ \top} \bZ)|T=0]\}^{-1} {E[\bZ\exp(\btheta^{ \top} \bZ)|T=0 ]}$. Let $ \mathbf{H}_j(\btheta) $ be the $ j $th row of $ \mathbf{H}(\btheta) $. Let $ A(\btheta)= \{E[\exp(\btheta^{ \top} \bZ)|T=0]\}^{-1} {E[Y\exp(\btheta^{ \top} \bZ)|T=0 ]} $. Let 	${\tau}_0$ be the true value of the ATT estimand, and ${\tau}^*$ be the asymptotic limit of the naive EB estimator $\hat{\tau}^*$.

The following theorem presents a  lower bound of  the asymptotic bias of the naive EB estimator $\hat{\boldsymbol{\theta}}^*$, and derives  bias formulas under the normal error assumption. Moreover, it characterizes  the asymptotic bias of the naive EB-based ATT estimator $\hat{\tau}^*$ using that of $\hat{\boldsymbol{\theta}}^*$. This theorem holds given that the EB estimator $\hat{\btheta}$ is consistent of ${\btheta}_0$.

	\begin{theorem}\label{thm2}
		(1). Under the  additive measurement error model \eqref{me},  the $L_q$ norm of the asymptotic bias of naive EB is lower bounded by
		\begin{equation}\label{biasbound}
\|\btheta^{*}-\btheta_0\|_q\geq \left\{
			\sup_{\bar{\btheta}\in L( \boldsymbol{\theta}^* ; \boldsymbol{\theta}_0)}\|\mathbf{H}(\bar{\btheta})\|_q\right\}^{-1}{ \|
			\nabla \log M(\btheta^*)\|_q}.
		\end{equation}

(2). Furthermore, if the  error term $\beps_{\scriptscriptstyle X}$ is  normally distributed,  then the following bias formula holds:
		\begin{equation}
			{\btheta}^{*}-\btheta_0 = -\left(
			\bar{\mathbf{H}} + \bSigma \right)^{-1}\bSigma \btheta_0,
			\label{biasformula}
		\end{equation}
		where $ \bar{\mathbf{H}}= (\mathbf{H}_1^\top(\bar{\btheta}_1), \cdots, \mathbf{H}_p^\top(\bar{\btheta}_p))^\top$ for some $\bar{\btheta}_j \in[\btheta_0,\btheta^{*}]$, $j=1,\cdots, p$. The  bias formula \eqref{biasformula} is equivalent to the following expression:
\begin{equation}\label{biasformula2}
			\begin{aligned}
				{\btheta}^{*}_{\bx} =& \left(\mathbf{I}_{p_1} +  \bar{\mathbf{H}}_{11}^{I}\bSigma_1 \right)^{-1} \btheta_{0,\bx},\\
				{\btheta}^{*}_{\bu} =&  \btheta_{0,\bu}-\bar{\mathbf{H}}_{21}^I\bSigma_1 \left(\mathbf{I}_{p_1}+\bar{\mathbf{H}}_{11}^{I}\bSigma_1 \right)^{-1} \btheta_{0,\bx}.
			\end{aligned}
		\end{equation}

(3). Moreover, the asymptotic bias of naive ATT estimation (${\tau}^*-{\tau}_0$)   and that of naive parameter estimation ($\btheta^{*}-\btheta_0$) have the following relationship:
		\begin{equation} \begin{aligned}
				{\tau}^*-{\tau}_0
				= -[\nabla A({\bar{\btheta}})]^\top(\btheta^{*}-\btheta_0),
			\end{aligned}\label{ATTbias}
		\end{equation}
		where $\bar{\btheta}\in[\btheta_0,\btheta^{*}]$.
	\end{theorem}

Assuming  that the errors are normally distributed and that $\btheta_{0,\bx}\neq \mathbf{0}_{p_1}$, we illustrate the consequences of Theorem \ref{thm2}. Noting that the assumption $\btheta_{0,\bx}\neq \mathbf{0}_{p_1}$ and equation \eqref{biasformula2} lead to that  ${\btheta}^{*}_{\bx} \neq \mathbf{0}_{p_1}$, we obtain that $\|\nabla \log M(\btheta^*)\|_q=\|\mathbf{\Sigma}_1\btheta_{\bx}^*\|_q>0$, and thus equation \eqref{biasbound}  implies that
$\|\btheta^{*}-\btheta_0\|_q>0$ and $\btheta^{*}\neq\btheta_0$.  Therefore, the naive EB estimator $\hat{\boldsymbol{\theta}}^*$ is inconsistent of $\btheta_0$ in the presence of measurement error.

The bias formula \eqref{biasformula}  characterizes the asymptotic bias ${\btheta}^{*}-\btheta_0$ and establish an explicit relationship of the bias and the measurement error magnitude.   To  illustrate the implication of this formula, consider a simple  structure of the  error variance-covariance matrix $\Sigma_1=\text{diag}(\sigma^2,\cdots, \sigma^2)$, where the error terms ${\epsilon}_{i,1},\cdots, {\epsilon}_{i,p_1}$ are independent and  the covariate data  are  rescaled such that the error variances ${\epsilon}_{i,j}$ are homoscedastic, $j=1,\cdots, p_1$.  Then, the bias formula \eqref{biasformula}  leads to that
$$\|{\btheta}^*\|_q\leq \left(\prod_{j=1}^{p_1}\frac{\lambda_j}
{\lambda_j+\sigma^2}\right)\|\btheta_0\|_q<\|\btheta_0\|_q,$$
where $\lambda_j>0$ is the $j$th largest eigenvalue of $\bar{\mathbf{H}}$, suggesting that the naive EB estimator underestimates the magnitude of the true value  $\btheta_0$ asymptotically. Moreover, the estimation bias  increases when the   measurement error variance increases. Therefore, the effect of measurement error amounts to attenuate naive EB estimation  to the null.

 The  second bias formula \eqref{biasformula2} delineates the relationship between ${\btheta}^{*}_{\bx}$ and $\btheta_{0,\bx}$ and the relationship between ${\btheta}^{*}_{\bu}$ and $\btheta_{0,\bu}$, respectively. It is similar to the bias analysis results for linear models \citep{Carroll}.   Except the special case that  $\bar{\mathbf{H}}_{21}^I=\mathbf{0}_{p_2\times p_1}$, the bias formula \eqref{biasformula2} implies that  naive estimation of $\btheta_{0,\bu}$ is biased, and this estimation bias tends to be larger when the error magnitude increases. Analogously, we obtain  that
$\|{\btheta}^{*}_{\bx}\|_q	 <\|\btheta_{0,\bx}\|_q$,
and  the estimation bias of parameter $\btheta_{\bx}$  generally increases when the magnitude of  measurement error increases. Using some matrix algebra, we obtain that the  asymptotic relative bias of naive estimation of $ {\btheta}_{\bx}$ is bounded to be
$$			\frac{1}{\|(\bar{\mathbf{H}}_{11}^{I})^{-1}\|_q\|\bSigma_1^{-1}\|_q+1}\leq
				\frac{\|\btheta^{*}_{\bx}-\btheta_{0,\bx}\|_q}{\|\btheta_{0,\bx}\|_q}\leq \frac{1}{\|(\bar{\mathbf{H}}_{11}^{I})^{-1}\|_q\|\bSigma_1^{-1}\|_q-1},
$$
		where   the upper bound holds only when $\|\bar{\mathbf{H}}_{11}^I\bSigma_1\|_q<1$.  The condition $\|\bar{\mathbf{H}}_{11}^I\bSigma_1\|_q<1$ is  met when the magnitude of the measurement error is small. As    the magnitude of measurement error decreases such that $\|\bSigma_1^{-1}\|_q$ increases, the upper and lower bounds   become tighter.

 The  third bias formula \eqref{ATTbias} shows that the asymptotic bias of naive ATT estimation is determined by the asymptotic bias $\btheta^{*}-\btheta_0$ and the gradient of the function $A({\btheta})$ at some intermediate value $\btheta=\bar{\btheta}$.  Because $\btheta^{*}-\btheta_0\neq \mathbf{0}_p$, it follows that
 ${\tau}^*-{\tau}_0\neq 0$
  in general, unless the gradient $\nabla A({\bar{\btheta}})$ is incidentally orthogonal to  $\btheta^{*}-\btheta_0$. Therefore, the naive EB-based ATT estimator $\hat{\tau}^*$ is inconsistent in most cases. In addition, the bias formula \eqref{ATTbias} immediately suggests an  upper bound: $|{\tau}^*-{\tau}_0|\leq {\|\nabla A({\bar{\btheta}})\|_{q^\prime}}\|\btheta^{*}-\btheta_0\|_q$,
where $\frac{1}{q^\prime}+\frac{1}{q}=1$.

Theorem \ref{thm2}  underscores the impact of measurement error under the EB framework,  highlighting the  biases of naive parameter and treatment effect estimation. When $\btheta_{0,\bx}\neq \mathbf{0}_{p_1}$, the presence of measurement error brings about persistent estimation bias that does not diminish even when the sample size increases to infinity. If $\btheta_{0,\bx}= \mathbf{0}_{p_1}$, then $\nabla \log M(\btheta_0)=\mathbf{0}_{p}$,  and thus equation \eqref{EB2} suggests that the naive EB method is consistent under the assumption of logistic propensity score model or linear outcome model.

\section{Covariate Balancing  with Measurement Error Correction \label{sec4}}

The bias analysis results in Section \ref{sec3} highlight the dramatic impact of measurement error on  covariate balance and parameter and treatment effect estimation. The claimed theoretical findings are confirmed  by the numerical studies presented in Section \ref{subsec5.1}. Therefore,   measurement error correction is  essential  if we attempt to restore covariate balance and achieve consistent estimation of treatment effects. In this section, we propose various correction strategies to eliminate the impact of measurement error.

In Section \ref{subsec4.0}, we discuss the general idea of restoring covariate balance under the EB or CBPS framework. In Sections \ref{subsec4.1},  \ref{sec4.2}, and \ref{subsec4.3}, we develop various methods for addressing measurement error under the EB framework. In specific,
 we assume that the measurement error distribution is either known or  consistently estimated in Section \ref{subsec4.1}.
In Section \ref{sec4.2}, we focus on handling measurement error by utilizing the bias formula   in Theorem \ref{thm2} under the normal error assumption.
In Section \ref{subsec4.3}, we relax the distributional error assumption   when replicated measurements of the error-prone covariates are available.
In  Section \ref{subsec4.4}, we explore  measurement error correction under the CBPS framework.

\subsection{Restoring covariate balance \label{subsec4.0}}
Under either the EB or the CBPS framework, consider  an arbitrary estimator of $\btheta$, named $\hat{\btheta}_c$,  which is tailored to correct for measurement error effect. This estimator utilizes the observed data set $\mathcal{O}_1^*$ and the distributional error assumption. In the following subsections, we propose various strategies to construct  $\hat{\btheta}_c$. For now, we assume that the estimator  $\hat{\btheta}_c$ is available.

Motivated by the weight construction in equation \eqref{EBweight} under the EB framework, it is natural to construct the corrected EB weight to be
\begin{equation}\label{cEBweight}
	\hat{w}_{c,i} = \frac{\exp({\hat{\btheta}}_c^{\top} \bZ_{i}^*)}{\sum\limits_{j:T_{j}=0} \exp({\hat{\btheta}}_c^{\top} \bZ_{j}^*)}
\end{equation}
for each subject $i$ in the control group.  Subsequently, we  propose the corrected EB estimator for the ATT:
  \begin{equation}
\hat{\tau}_c=\bar{Y}_{1} - \sum_{i:T_i=0} \hat{w}_{c,i} Y_{i}.\label{cATT}
\end{equation}
Moreover, motivated by the  weight formulation \eqref{CBPSweight},  it is natural to construct the  corrected CBPS weight to be
 \begin{equation}\label{cCBPSweight}
\hat{w}_{c,i} = \frac{\pi_{\hat{\btheta}_c}(\bZ_i^*)}{n_1\{1-\pi_{\hat{\btheta}_c}(\bZ_i^*)\}}
\end{equation}
for subject $i$ in the control group. We then  estimate the ATT accordingly.

To evaluate the performance of the aforementioned corrected  methods for balancing $ \bZ $, we use the corrected ASMD imbalance measure:
 \begin{equation}\label{cASMD}
{ASMD}_{c,j}=\frac{\left|\bar{Z}_{j}^{trt}-\sum\limits_{i:T_i=0} \hat{w}_{c,i}  Z_{ij}\right|}{\hat{\sigma}_{j}^{trt}},  \ \ j=1, \cdots, p,
\end{equation}
where $\hat{w}_{c,i}$ is the corrected EB  weight in equation \eqref{cEBweight} or the  corrected CBPS weight in equation \eqref{cCBPSweight}. Similarly, define ${MD}_c$ to be ${MD}$ with $\hat{w}_{i}$ replaced by $\hat{w}_{c,i}$.

The following theorem asserts that if  $\hat{\btheta}_c$ is   consistent of ${\btheta}_0$, then \textit{large-sample  balance} holds simultaneously for the unobserved true covariates $\bX$ and the precisely measured covariates $\bU$ under the EB or CBPS framework. This theorem holds without assuming the  logistic propensity score model or the linear outcome model.

\begin{theorem}\label{prop1}
 Assume that the additive measurement error model holds. Suppose that $\hat{\btheta}_c$ is a consistent estimator of ${\btheta}_0$. As $n\rightarrow \infty$,
	\[
	\begin{aligned} \text{ASMD}_{c,j}&\stackrel{p}{\longrightarrow}
0,\ j=1,\cdots, p,\\
\text{MD}_c&\stackrel{p}{\longrightarrow}{0}.
	\end{aligned}
	\]
\end{theorem}

Theorem \ref{prop1} addresses the importance of adequately  estimating the parameter ${\btheta}_0$. This problem is resolved  in Sections \ref{subsec4.1}-\ref{subsec4.4}, where we propose various measurement error correction methods to obtain consistent or approximately consistent estimators of ${\btheta}_0$.

We remark that it is impossible to  fully recover \textit{finite-sample  balance} for all components of the underlying true covariates $ \bZ=(\bX^\top,\bU^\top)^\top$ because the $\bX$ are  latent. However, we show  in the  following subsections that it is possible to maintain  \textit{finite-sample  balance} for  the precisely measured covariates $\bU$ and meanwhile achieve  \textit{large-sample  balance} for  the  unobservable true covariates $\bX$.

\subsection{Corrected  entropy balancing: a consistent measurement error correction approach \label{subsec4.1}}

Inspired by Theorem \ref{prop1}, we aim to  construct a consistent estimator $\hat{\btheta}_c$ of ${\btheta}_0$. We focus on the EB framework in Sections \ref{subsec4.1}-\ref{subsec4.3},
whereas  we explore the CBPS framework in  Section \ref{subsec4.4}.

Recall that  Lemma \ref{lem1} shows that the effect of measurement error on naive EB estimation is to introduce the  regularization term $ \log M(\btheta)$ to the  loss function $L(\btheta; \mathcal{O}_1)$  in equation \eqref{EB2}. To compensate the regularization effect, it is natural to subtract the term  $ \log M(\btheta)$ from the loss function $L(\btheta; \mathcal{O}_1^*)$. We refer to this strategy as corrected entropy balancing (CEB). The proposed CEB estimator of $\btheta$ is:
\begin{equation}
	\hat{\btheta}_c=\underset{{{\btheta}\in \mathbb{R}^p}}{\text{argmin}}  \
	\left\{L(\btheta; \mathcal{O}_1^*)-\log M(\btheta)\right\}.\label{CEB}
\end{equation}
The  CEB weight $\hat{w}_{c,i}$ is obtained from equation \eqref{cEBweight} for subject $i$ in the control group. Subsequently, the CEB-based ATT estimator $\hat{\tau}_c$ is obtained from equation \eqref{cATT}.
When the errors are normally distributed, the CEB estimator is
\begin{equation*}
	\hat{\btheta}_c=\underset{{\btheta}\in \mathbb{R}^p}{\text{argmin}}\left\{ L(\btheta; \mathcal{O}_1^*)- \frac{1}{2}\btheta_{\bx}^\top{\bSigma}_1\btheta_{\bx}\right\},
\end{equation*}
where the term $\frac{1}{2}\btheta_{\bx}^\top{\bSigma}_1\btheta_{\bx}$ serves to counteract the shrinkage  effect of measurement error.
Interestingly, we show that the precisely measured covariates $\bU$ are exactly balanced using the CEB method. This theorem holds without assuming the  logistic propensity score model or the linear outcome model.
\begin{theorem}
	\label{cor2}
		Assume that the additive measurement error model holds.   Finite-sample balance  holds  for  $\bU$ by the CEB method in the sense that
	\[
	\begin{aligned} \text{ASMD}_{c,j}&=
		0,\ j=p_1+1,\cdots, p.
	\end{aligned}
	\]
\end{theorem}

In the following theorem, we show that the CEB estimator $\hat{\btheta}_c$ is consistent. Therefore, the results in Theorems \ref{prop1} are applicable to  the CEB method.  In conclusion, the proposed CEB method   maintains \textit{finite-sample balance} for     $\bU$ and meanwhile it assures \textit{large-sample balance} for  $\bX$.

Note that the gradient of the loss function in equation \eqref{CEB} is   asymptotically unbiased of $\nabla L(\btheta; \mathcal{O}_1)$ conditional on the underlying data $\mathcal{O}_1$:
\begin{equation}
	E\left[\left.\nabla L(\btheta; \mathcal{O}_1^*)-\frac{\nabla M(\btheta)}{M(\btheta)}\ \right|\  \mathcal{O}_1 \right]=\nabla L(\btheta; \mathcal{O}_1)+o_p(1).\label{CEBEE}
\end{equation}
By the theory of estimating equations and the property \eqref{CEBEE}, we derive the consistency and asymptotic normality results  in the following theorem. The proofs and regularity conditions are relegated to the Supplementary Material.

	\begin{theorem}\label{thm:AsymptoticTheory}
	Assume  the logistic propensity score model $\pi_{{\btheta}}(\bZ)=\text{logit}^{-1}(\bZ^\top \btheta)$.  As $n\rightarrow\infty$, $\hat{\btheta}_c\xrightarrow{p} \btheta_0$ and $\hat{\tau}\xrightarrow{p}\tau_{0}$. Furthermore, as $n\rightarrow\infty$,
		\[
		\begin{aligned}
			&\sqrt{n} (\hat{\btheta}_c - \btheta_{0}) \xrightarrow{d} {MVN}(\mathbf{0}_p,\mathbf{\Sigma}_{\btheta_0}),\\
			&\sqrt{n}(\hat{\tau}_c - \tau_{0})\xrightarrow{d} {N}(0,\Sigma_{\tau_0}),
		\end{aligned}
		\]
where the variance-covariance matrices $\mathbf{\Sigma}_{\btheta_0}$ and $\Sigma_{\tau_0}$ are defined in  the Supplementary Material.
	\end{theorem}

In addition to consistency and asymptotic normality, the CEB-based ATT estimator $\hat{\tau}_c$ is doubly robust. This property is crucial when either the propensity score model or the outcome model is uncertain.

	\begin{theorem}
		\label{thm:dr}
	 Assume that the additive measurement error model holds. The CEB-based ATT estimator $\hat{\tau}_c$ is doubly robust  in the sense that if either the logistic propensity score model or the linear outcome model is correct, then $\hat{\tau}_c$  achieves consistency.
	\end{theorem}

\subsection{Bias-corrected  entropy balancing: a simple approximation approach}\label{sec4.2}

In the subsection, we propose a bias-corrected entropy balancing (BCEB) estimator of $\btheta$ when the errors are normal and the variance-covariance matrix $\bSigma$ is either known or consistently estimated. The BCEB estimator adjusts for the bias of the naive estimator $\hat{\btheta}^*$ via the bias formula \eqref{biasformula} as follows. By inverting   the bias formula \eqref{biasformula}, we obtain that
		\begin{equation}
			\btheta_0 = \bar{\mathbf{H}}^{-1}\left(
			\bar{\mathbf{H}} + \bSigma \right) {\btheta}^{*}.
			\label{biasformula3}
		\end{equation}
To obtain a good estimator of $\btheta_0$, equation \eqref{biasformula3} suggests that we need to   estimate $\bar{\mathbf{H}}$ and  ${\btheta}^{*}$ reasonably well. Recall that the naive estimator $\hat{\btheta}^*$ converges in probability to ${\btheta}^{*}$ as $n\rightarrow\infty$. It remains to construct a good estimator of $\bar{\mathbf{H}}$.

It can be shown that $\nabla^2 L(\btheta; \mathcal{O}_1^*)\stackrel{p}{\longrightarrow} \mathbf{H}(\btheta)+\bSigma$ {as} $n\rightarrow\infty$. Therefore, we consistently approximate $\mathbf{H}(\btheta)$ by $\nabla^2 L(\btheta; \mathcal{O}_1^*)-\bSigma$  for any fixed $\btheta$ in its parameter space. Nevertheless, recall that $ \bar{\mathbf{H}}= (\mathbf{H}_1(\bar{\btheta}_1)^\top, \cdots, \mathbf{H}_p(\bar{\btheta}_p)^\top)^\top$ where $\bar{\btheta}_j \in[\btheta_0,\btheta^{*}]$ is unknown, $j=1,\cdots, p$. We approximate $\bar{\btheta}_j$ by $\hat{\btheta}^*$.  Therefore,  $\bar{\mathbf{H}}$  is approximated by $\mathbf{H}(\hat{\btheta}^*)$, which is further approximated by $\nabla^2 L(\hat{\btheta}^*; \mathcal{O}_1^*)-\bSigma$.

Replacing  $\bar{\mathbf{H}}$ and ${\btheta}^{*}$ with $\nabla^2 L(\hat{\btheta}^*; \mathcal{O}_1^*)-\bSigma$ and  $\hat{\btheta}^*$   in equation \eqref{biasformula3} respectively,  we obtain the  BCEB estimator of $\btheta$:
\begin{equation}
\hat{\btheta}_{c}=\left\{
		\nabla^2 L(\hat{\btheta}^*; \mathcal{O}_1^*)-\bSigma \right\}^{-1} \left\{ \nabla^2 L(\hat{\btheta}^*; \mathcal{O}_1^*) \right\}\hat{\btheta}^*.\label{BCEB}
\end{equation}
Subsequently, we construct the  BCEB weights
and the BCEB-based ATT estimator using equations \eqref{cEBweight} and \eqref{cATT}, respectively.

Although the estimator $\hat{\btheta}_{c}$ in equation \eqref{BCEB} is generally inconsistent of $\btheta_0$ because $\bar{\btheta}_j\neq {\btheta}^*$, the numerical studies suggest that the biases of the BCEB and CEB estimators are  similar under the normal error assumption. The BCEB method demonstrates greater numerical stability compared to the CEB method due to its simple structure.  It  performs reasonably well  under non-normal error distributions, despite the fact that the bias formula relies on the normality assumption.

In the following subsection, we  show that  the distributional assumption of measurement error can be removed when replicated measurements are available.

\subsection{Corrected entropy balancing with replicated measurements: a distributional assumption-free scheme \label{subsec4.3}}

In this subsection, we explore the scenario where each subject $i$ has $m_i \ (m_{i} \geq 1)$ replicated measurements of covariates, denoted as $\bZ_{ij}^{*}$ for $j=1,\cdots, m_i$ \citep{Yan2}. Analogous to  the additive error model \eqref{me}, we assume that
\begin{equation}
	\bZ_{ij}^{*} = \bZ_{i} + \beps_{ij}, \label{me2}
\end{equation}
where the set $\{ \beps_{ij}: i=1,\cdots,n; j = 1,\cdots,m_i \}$ comprises independent and identically distributed errors with zero mean, and they are independent of the true data set $\mathcal{O}$. We make no distributional assumption for these error terms. Define the data set $\mathcal{O}_1^{**} = \{(T_{i}, {\bZ}_{ij}^{*}): i=1, \cdots, n, j=1,\cdots, m_i\}$.

A notable usage of replicated measurements in measurement error analysis \citep{Carroll} is that the variance-covariance matrix  $\bSigma$ is consistently estimated by
$$\hat{\bSigma}=\frac{\sum_{i=1}^{n} \sum_{j=1}^{m_{i}}\left(\bZ^{*}_{ij}-\bar{\bZ}^{*}_{i\cdot}\right)^{\otimes 2}}{\sum_{i=1}^{n}\left(m_{i}-1\right)},$$
where  $\bar{\bZ}^{*}_{i\cdot} = \sum_{j=1}^{m_{i}} \bZ_{ij}^{*} / m_{i}$, and $ \mathbf{a}^{\otimes 2} = \mathbf{a}\mathbf{a}^{\top} $ for any vector $ \mathbf{a} $.

Analogous to equation \eqref{CEBEE}, we aim to construct an asymptotically unbiased estimating function $\mathbf{B}(\btheta; \mathcal{O}_1^{**})$ such that
\begin{equation}
	\label{EE.rep}
	E\left[\mathbf{B}(\btheta; \mathcal{O}_1^{**}) \ |\  \mathcal{O}_1 \right] = \nabla L(\btheta; \mathcal{O}_1) +o_p(1)
\end{equation}
 holds for any $\btheta$ in its parameter space. The proposed corrected  estimator $\hat{\btheta}_{c}$ is obtained by solving
 $$\mathbf{B}(\btheta; \mathcal{O}_1^{**})=\mathbf{0}_p.$$
Using the theory of estimating equations and the property \eqref{EE.rep}, this estimator is guaranteed to be consistent of  $\btheta_0$.

We first impose a symmetric  distribution assumption that $\beps_{ij}$ and $-\beps_{ij}$ share the same distribution. Define $\bet_k(\btheta) = E\{ \exp(\btheta^{\top} \beps_{ij})\beps_{ij}^{\otimes k} \}$ for $k = 0,1$. Let
$R_{i}^{(0)}(\btheta) = {m_{i}^{-1} \eta_0^{-1}(\btheta)}{\sum_{j=1}^{m_{i}} \exp (\btheta^{\top} \bZ_{ij}^{*})}$  and	$\mathbf{R}_{i}^{(1)}(\btheta) = R_{i}^{(0)}(\btheta)  \{\bZ_{ij}^{*} - {\eta_{0}^{-1}(\btheta)}{\bet_{1}(\btheta)}\}$. It is easy to verify that $E[R_{i}^{(0)}(\btheta) | \mathbf{Z}_{i}]
=\exp (\btheta^{\top} \bZ_{i})$ and
$E[\mathbf{R}_{i}^{(1)}(\btheta) | \mathbf{Z}_{i}]
= \exp(\btheta^{\top} \bZ_{i})\bZ_{i}$.
Because $E[\exp \{\btheta^{\top}(\mathbf{Z}_{ij}^*-\mathbf{Z}_{ik}^*)\}] = \eta_{0}^{2}(\btheta)$ due to the symmetric assumption,  $\eta_{0}(\btheta)$ is consistently approximated by
\[
\hat{\eta}_{0}(\btheta) = \left[\left(\sum_{j=1}^{n} \xi_{j}\right)^{-1}\sum_{i=1}^{n}\left\{ \frac{\xi_{i}}{m_{i}(m_{i}-1)} \sum_{1 \leq j \neq k \leq m_i} \exp \left(\btheta^{\top}\left(\bZ_{ij}^{*}-\bZ_{ik}^{*}\right)\right)\right\}\right]^{1/2},
\]
where $\xi_{i} = I(m_{i} > 1)$ indicates the inclusion of subjects with more than one measurement, and the term $\xi_{i}/(m_{i}-1)$  is defined to be zero when $m_i=1$. Define $\hat{\bet}_{1}(\btheta)=\nabla \hat{\eta}_{0}(\btheta)$. Let $\hat{R}_{i}^{(0)}(\btheta)$ and $\hat{\mathbf{R}}_{i}^{(1)}(\btheta)$ be  $R_{i}^{(0)}(\btheta)$ and $\mathbf{R}_{i}^{(1)}(\btheta)$  where ${\eta}_{0}(\btheta)$ and ${\bet}_{1}(\btheta)$ are replaced by $\hat{\eta}_{0}(\btheta)$ and $\hat{\bet}_{1}(\btheta)$, respectively.

Inspired by the estimating function strategy in \cite{hu2004semiparametric}, we propose the following  estimating function:
$$
\mathbf{B}_{HL}(\btheta; \mathcal{O}_1^{**}) =\frac{\sum\limits_{i:T_i=0}\hat{\mathbf{R}}_{i}^{(1)}(\btheta)}
{\sum\limits_{i:T_i=0}\hat{R}_{i}^{(0)}(\btheta)}-\frac{\sum\limits_{i:T_i=1}\sum\limits_{j = 1}^{m_i}\bZ_{ij}^{*}}{\sum\limits_{i:T_i=1}  m_{i}}.
$$
It can be shown that this function $\mathbf{B}_{HL}(\btheta; \mathcal{O}_1^{**})$ possesses  the asymptotic unbiasedness property \eqref{EE.rep}  \citep{hu2004semiparametric}. We propose the CEB-HL estimator $\hat{\btheta}_{c}$ by solving
$$\mathbf{B}_{HL}(\btheta; \mathcal{O}_1^{**}) = \mathbf{0}_p.$$

Moving ahead, we remove the assumption of symmetric error distribution.
Motivated  by the nonparametric correction  strategy in \cite{huang2000cox}, we propose to use
the following estimating function:
$$\mathbf{B}_{HW}(\btheta; \mathcal{O}_1^{**}) =\frac{\sum\limits_{i:T_i=0} \left\{m_{i}^{-1} (m_{i}-1)^{-1}\xi_i\sum\limits_{ 1\leq j \neq k\leq m_i}\exp(\btheta^{\top} \bZ_{ij}^{*})\bZ_{ik}^{*}\right\}}{ \sum\limits_{i:T_i=0} \left\{ m_{i}^{-1}\xi_i\sum\limits_{j=1}^{m_{i}}\exp(\btheta^{\top}\bZ_{ij}^{*})\right\}}
-\frac{\sum\limits_{i:T_i=1} \bar{\bZ}^{*}_{i\cdot}}{n_{1}}.$$
It can be shown that this function $\mathbf{B}_{HW}(\btheta; \mathcal{O}_1^{**})$ possesses  the asymptotic unbiasedness property \eqref{EE.rep}  \citep{huang2000cox}. We propose the CEB-HW estimator $\hat{\btheta}_{c}$  by solving $$\mathbf{B}_{HW}(\btheta; \mathcal{O}_1^{**})=\mathbf{0}_p.$$

 Next, we discuss the    construction of weights for the CEB-HL and  CEB-HW methods. In the presence of replicated measurements, the weight formula in equation \eqref{cEBweight} is no longer suitable. For subject $i$ in the control group, we propose the following  corrected weight:
\begin{equation}\label{Replicatedweight}
\hat{w}_{c,i} = \frac{
 {m_{i}^{-1}}\sum\limits_{j=1}^{m_{i}} \exp \left(\hat{\btheta}_c^{\top} \bZ_{ij}^{*}\right)
}{ \sum\limits_{l:T_l = 0} {m_{l}^{-1}}{\sum\limits_{j=1}^{m_{l}} \exp \left(\hat{\btheta}_c^{\top} \bZ_{lj}^{*}\right)}
}.
\end{equation}
The   CEB-HW or CEB-HL-based ATT estimator can then be obtained via equation \eqref{cATT} using the weights in \eqref{Replicatedweight}.

Interestingly, similar to the CEB method in Section \ref{subsec4.1},  the precisely measured covariates $\bU$ are exactly balanced  by the CEB-HL or CEB-HW method. This theorem holds without assuming the  logistic propensity score model or the linear outcome model.

\begin{theorem}
	\label{cor22}
		Assume that the additive measurement error model holds.  Finite-sample balance holds  for  $\bU$ when the CEB-HL weights  are used   in  equation \eqref{cASMD}:
	\[
	\begin{aligned} \text{ASMD}_{c,j}&=
		0,\ j=p_1+1,\cdots, p.
	\end{aligned}
	\]
Moreover, if $m_i>1$ for all subjects in the control group, then finite-sample balance holds  for  $\bU$ when the CEB-HW weights  are used   in  equation \eqref{cASMD}.
\end{theorem}

The consistency and asymptotic normality of the CEB-HL and CEB-HW methods can be established using  asymptotic theory. Nevertheless, the underlying theoretical arguments  are  complicated, which are beyond the focus of this paper. As such, a detailed theoretical exploration  is not presented in this work.

\subsection{Corrected covariate balancing propensity score \label{subsec4.4}}

In this subsection, we discuss measurement error correction under the CBPS framework.
We employ the logistic propensity score model and assume  that  the errors are normally distributed.

Recall that there are two sets of estimating equations in equation \eqref{eq:cbpsbalancing}, where the first set consists of the likelihood score equations, and the second set consists of  the balancing conditions. We aim to correct for measurement error for  the likelihood score equations and the balancing conditions, respectively.

In the presence of measurement error, we propose to replace  the likelihood score equations with the conditional score equations:
\begin{equation}
	\sum_{i=1}^n \left[T_i-\frac{1}{1+\exp\left\{-\btheta^\top \bZ_i^*-(T_i-0.5)\btheta_{\bx}^\top {\bSigma}_1\btheta_{\bx}\right\}}\right]\bZ_i^*=\mathbf{0}_p.\label{cscore}
\end{equation}
The solution of the conditional score equations  \eqref{cscore} is consistent of $\btheta_0$ \citep{StefanskiCarroll1987}.

Next, we discuss the balancing conditions. Under the logistic propensity score model, the  balancing conditions is
\begin{equation*}
	\sum_{i:T_i=0}w_i\bZ_i-\bar{\bZ}^{\text{trt}}=\mathbf{0}_p,
\end{equation*}
where $w_i=n_1^{-1} \exp(\btheta^\top{\bZ_i})$.

Analogous to the  estimating equation-based measurement error correction strategies in equations \eqref{CEBEE} and \eqref{EE.rep},  we aim to construct an estimating function $\mathbf{C}(\btheta; \mathcal{O}_1^{*})$ that satisfies the property:
\begin{equation}\label{cbpsEE}
	E\left[\mathbf{C}(\btheta; \mathcal{O}_1^{*}) \ |\  \mathcal{O}_1 \right]=n_1^{-1}\sum_{i:T_i=0} {\bZ_i\exp(\btheta^\top{\bZ_i}) }-\bar{\bZ}^{trt}+o_p(1).
\end{equation}
It can be shown that the estimating function
\begin{equation*}
	\mathbf{C}_{HW}(\btheta; \mathcal{O}_1^{*})=
	n_1^{-1}\sum_{i:T_i=0} \left(\bZ_i^*-{\bSigma}\btheta\right)\exp\left(\btheta^\top \bZ_i^*-0.5\btheta_{\bx}^\top{\bSigma}_1\btheta_{\bx}\right)-\bar{\bZ}^{*,trt}
\end{equation*}
satisfies the property \eqref{cbpsEE}. Interestingly,  this estimating function coincides with  equation (8) in \cite{HuangWang2001}, which was tailored for logistic regression with measurement error. The solution of
\begin{equation}
\mathbf{C}_{HW}(\btheta; \mathcal{O}_1^{*})=\mathbf{0}_p\label{Bhw}
 \end{equation}
is consistent of $\btheta_0$ \citep{HuangWang2001}.

We propose the corrected CBPS  estimator $\hat{\btheta}_c$ by solving the   conditional score equations
\eqref{cscore} and the corrected balancing conditions \eqref{Bhw}
 simultaneously. Either the generalized method of moments  or the empirical likelihood method can be employed \citep{imai2014covariate}.

We construct the corrected  CBPS weight using equation \eqref{cCBPSweight} and obtain that
\begin{equation}\label{normalizedcCBPSweight}
	\hat{w}_{c,i} =\frac{\exp({\hat{\btheta}}_c^{\top} \bZ_{i}^*)}{\sum\limits_{j:T_{j}=0} \exp({\hat{\btheta}}_c^{\top} \bZ_{j}^*)}.
\end{equation}
 The corrected CBPS-based ATT estimator is   $\hat{\tau}_c=\bar{Y}^{trt} - \sum_{i: T_i=0} \hat{w}_{c,i} Y_{i}$. We remark that the corrected CBPS weights in equation \eqref{normalizedcCBPSweight} have the same expression as the corrected EB weights in equation \eqref{cEBweight}.

If  replicated measurements are available, it is feasible to construct corrected balancing conditions using the functional correction strategy \citep{HuangWang2001} which is free of distributional error assumption. Further investigation of corrected CBPS methods is warranted.

\section{Numerical Studies \label{sec5}}

This section is organized as follows. In Section \ref{subsec5.1}, we present a numerical study  assessing the effect of measurement error on the naive EB method. In Section \ref{sec5.2}, we undertake a comparative evaluation of naive covariate balancing against the proposed corrected balancing methods  in Section \ref{sec4} in estimating the ATT. In Section \ref{sec5.2.2}, we compare  these methods in balancing covariates. In Section \ref{sec5.3}, we  conduct real data analysis.

\subsection{Impact of measurement error \label{subsec5.1}}

We conduct a large-sample numerical study for naive EB, where we investigate the impact of measurement error on covariate balance, parameter estimation,  and treatment effect estimation. We run 1000 Monte Carlo simulations, each generating a sample of size $n=50000$.

The simulation setup is similar to that  in \cite{zhao2016entropy}. In specific,
the covariate $\bZ=(X_1, U_1)^\top$ is generated from a bivariate normal distribution with mean vector $(5,10)$, variance $\text{Var}(X_1)=\text{Var}(U_1)=1$, and covariance $\text{Cov}(X_1,U_1)=0.3$.
Given the covariate $\bZ$, the binary treatment indicator $T$ is simulated from a Bernoulli distribution where the success probability $P(T=1 | \bZ)$ follows the logistic propensity score model: $\text{logit}(P(T=1 | \bZ)) = 0.5 - 3X_1 + 1.5 U_1$.
According to the relationship between EB and logistic regression, the true parameter value in the dual problem \eqref{eq:dual-opt-EB} of EB is $\btheta_0 = (-3, 1.5)^\top$. The observed outcome is obtained by $Y=TY(1)+(1-T)Y(0)$, where the potential outcomes are generated as follows:   $Y(1)\sim N(220+27.4X_1+13.7U_1,4)$, $Y(0)\sim N(210+27.4X_1+13.7U_1,4)$, and $Y(1)$ and $Y(0)$ are generated independently given $(X_1,U_1)$.
The true ATT value is $\tau_0 = 10$.

Suppose that the true covariate $X_1$ is obscured by measurement error and thus is unobserved. Instead, we observe $X_1^* \sim N(X_1, \sigma_1^2)$, where  $\sigma_1^2\in\{0.00, 0.05, 0.10, \ldots, 0.50\}$. Accordingly,  the noise-to-signal ratio $\sigma_1^2/\text{Var}(X_1)$ ranges  from 0\% (indicating no measurement error) to 50\% (indicating severe
  measurement error).

\begin{figure}[h]
	\centering \includegraphics[scale=0.50]{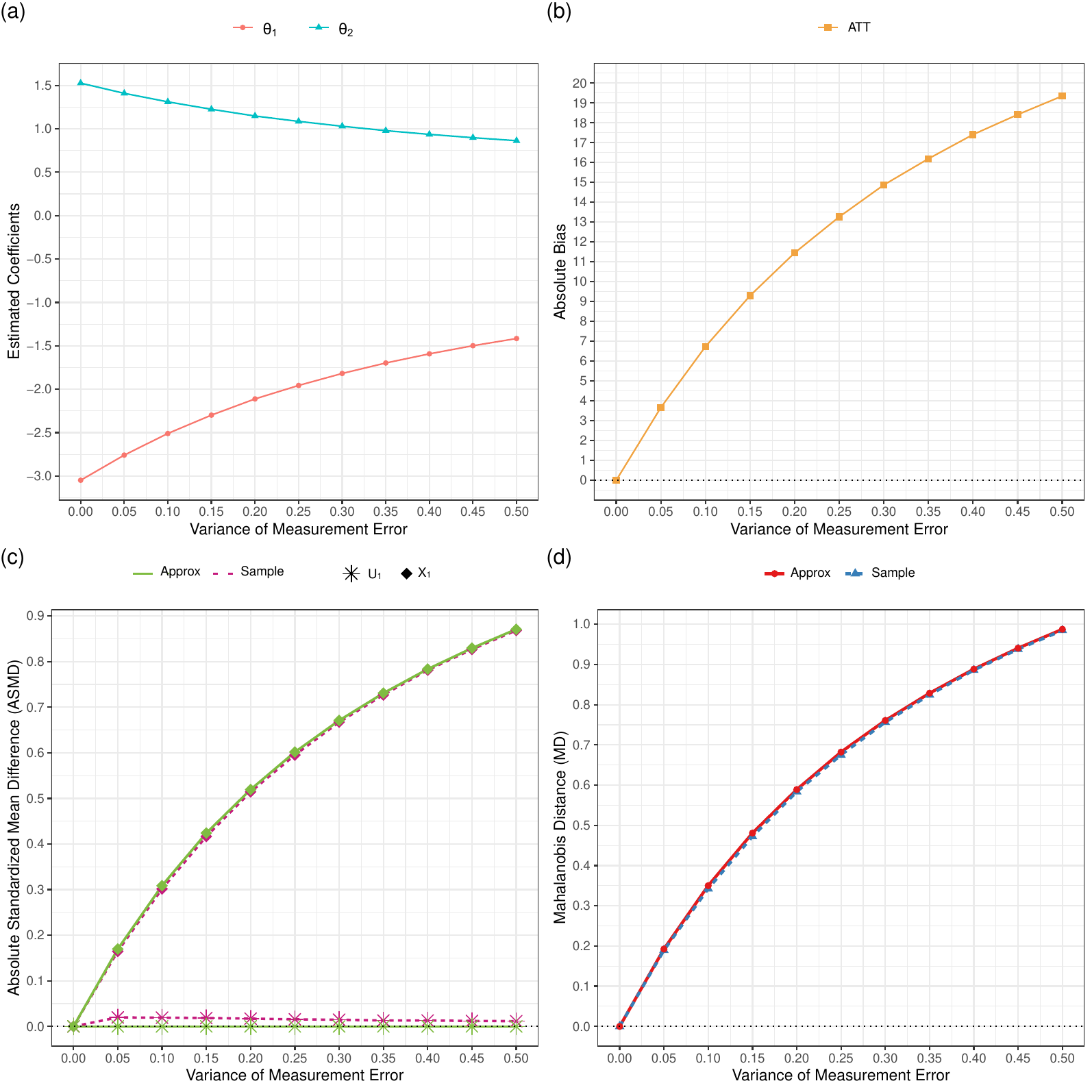}
	\caption{Impact of measurement error on the naive EB method. Plot (a) shows the naive EB estimator $\hat{\btheta}^*=(\hat{\theta}^*_1,\hat{\theta}^*_2)^\top$. The true value is $\btheta_0=(-3.0, 1.5)^\top$; Plot (b) shows the average  absolute bias of naive EB-based ATT estimation: $|\hat{\tau}^*-\tau_0|$; Plot (c) depicts the ASMD metrics ${ASMD}_j^*$ for  $X_1$ and $U_1$, $j=1,2$,  calculated by the definition \eqref{nEBASMD} using the simulated samples (Sample). We also report the approximated version of ${ASMD}_j^*$ obtained by the asymptotic limit   in Theorem \ref{thm1} (Approx); Plot (d) depicts  the MD metrics  calculated by the definition (Sample) and by asymptotic approximation (Approx).}
	\label{pBA_2Covs}
\end{figure}

We apply naive EB  in Section \ref{subsec2.4} to estimate the parameter $\btheta$ and the ATT $\tau$. Figure \ref{pBA_2Covs} summarizes the simulation outputs. In Plot (a) of Figure \ref{pBA_2Covs}, we report the Monte Carlo average of the naive EB estimator $\hat{\btheta}^*=(\hat{\theta}^*_1,\hat{\theta}^*_2)^\top$ obtained by solving the dual problem \eqref{naiveEB}. Plot (a) shows that as measurement error increases, the naive EB estimators $(\hat{\theta}^*_1,\hat{\theta}^*_2)$ increasingly deviate from their true values $(-3,1.5)$, and they are shrunk towards zero. Therefore, Plot (a) demonstrates the attenuation phenomenon asserted by Theorem \ref{thm2}. This attenuation effect is amplified when the noise-to-signal ratio approaches  50\%.

In Plot (b), we report the   average  of the absolute bias of the naive EB-based ATT estimator $|\hat{\tau}^*-\tau_0|$. Plot (b) displays  an evident upward trajectory, which is markedly more pronounced than the bias of  parameter estimation  observed in Plot (a). This disparity stems from the influence of the negative gradient $-\nabla A({\bar{\btheta}})$ in  the bias formula \eqref{ATTbias} in Theorem \ref{thm1}, which  accelerates the bias trend in Plot (b). The inferior performance of the naive EB method in parameter and treatment effect estimation is attributable to the imbalance of the true covariate $X_1$, as depicted in Plots (c) and (d).

In Plot (c),  we  show  the Monte Carlo average ASMD values for $X_1$ and  $U_1$ obtained by equation \eqref{nEBASMD}, where   the naive EB weights are employed.  For comparison purpose, we report the asymptotic limit of the ASMD  metric  by Theorem \ref{thm1}. Plot (c) shows close alignment between the Monte Carlo average (Sample) and asymptotic (Approx) versions of  the  ASMD   values, thereby empirically validating   Theorem \ref{thm1}. The ASMD metric for $X_1$ demonstrates an evident upward trend as the variance of measurement error $\sigma_1^2$ increases. This might appear counterintuitive at first glance since   $\hat{\theta}^*_1$ and $\hat{\theta}^*_2$ are attenuated to the null  in Plot (a). However, the upward trend in Plot (c) can be explained by Theorem \ref{thm1}. According to Theorem \ref{thm1}, the asymptotic limit of  $A S M D_1^*$  is $\sigma_1^2|{\theta}^*_1|/\sigma_j^{\text {trt }}$, where $\sigma_j^{\text {trt }}\approx 0.8$ in this simulation setup. Therefore, the  $A S M D_1^*$ value in Plot (c) can be  calculated using Plot (a): $\sigma_1^2|{\theta}^*_1|$ increases from 0 to about 0.7 when $\sigma_1^2$ increases from 0 to 0.50, and hence $A S M D_1^*\approx\sigma_1^2|{\theta}^*_1|/0.8$ increases from 0 to about 0.88.

The threshold 0.20 or 0.25 is commonly used for the ASMD metric, and an ASMD value greater than the threshold indicates that the corresponding covariate is imbalanced.  In this manner, Plot (c) suggests that the true covariate $X_1$ becomes considerably imbalanced  even when the magnitude of measurement error is  small: $A S M D_1^*>0.25$ when the noise-to-signal ratio exceeds merely $8\%$.  In contrast, the precisely measured covariate $U_1$ is well balanced even when the noise-to-signal ratio is 50\%. This confirms the finding in Theorem \ref{thm1}.

Analogously, we display the empirical (Sample) and asymptotic (Approx) versions of  the   MD  metric in Plot (d).  The two curves closely align with each other, both displaying an rapidly increasing trend of overall residual imbalance.

In conclusion, measurement error has a detrimental impact on  naive EB. It distorts covariate balance and introduces substantial bias to naive ATT estimation. Numerical studies not reported in this paper show similar  measurement error effect on   naive CBPS and other covariate balancing methods.   Therefore, it is necessary to adjust for any existing covariate balancing methods to mitigate the negative impact of measurement error on covariate balance and  on the validity of causal inference.

\subsection{ATT estimation with measurement error correction\label{sec5.2}}

In this subsection, the performance of naive EB is evaluated against the correction methods outlined in Section \ref{sec4}: CEB, BCEB, CEB-HL, and CEB-HW. We focus on ATT estimation.  Numerical results for the corrected CBPS are documented elsewhere.

We conduct 1000 Monte Carlo simulation runs  for each parameter configuration. The sample size is $ n = 2000 $. We create four  variables $\bZ=(X_1,X_2,U_1,U_2)^\top$ from a   multivariate normal distribution. Given the covariate $\bZ$, the binary treatment indicator $T$ is simulated from a Bernoulli distribution where $\text{logit}(P(T=1 | \bZ)) = 3.5-X_1+0.5X_2-0.25 U_1-0.1U_2$.
The observed outcome is obtained by $Y=TY(1)+(1-T)Y(0)$, where   $Y(1)\sim N(220+27.4X_1+13.7X_2+13.7U_1+13.7U_2,4)$ and $Y(0)\sim N(210+27.4X_1+13.7X_2+13.7U_1+13.7U_2,4)$, and $Y(1)$ and $Y(0)$ are independent given $\bZ$. The true ATT value is $\tau_0 = 10$.

The covariates $\bX=(X_1,X_2)^\top$ are subject to measurement error, whereas  $\bU=(U_1,U_2)^\top$  are precisely measured. The mismeasured covariates are generated from the additive measurement error model  \eqref{me2} with $m_i=2$ for $i=1,\cdots, n$: $
\bZ_{ij}^{*} = \bZ_{i} + \boldsymbol{\epsilon}_{ij}$, where $\bZ_{ij}^{*}=(X_{1ij},X_{2ij}, U_{1i}, U_{2i})^\top$
 and
$\boldsymbol{\epsilon}_{ij}=({\epsilon}_{1ij},{\epsilon}_{2ij},0,0)^\top$.
The two error terms ${\epsilon}_{1ij}$ and ${\epsilon}_{2ij}$ are set to be independent and identically distributed. We consider four cases for the error distribution: normal distribution $N(0,\sigma_{1}^{2})$ (Case I),  uniform distribution (Case II), modified Beta distribution $ \text{Beta}(3,1) $ (Case III) or scaled $t$ distribution $t_3$ (Case IV). The mean of the errors is zero and the variance is set to be 0.1 or 0.5.  The Beta distribution is modified and the $t$ distribution is scaled to satisfy the predetermined mean and variance values.  These cases allow us to obtain an informative  evaluation of  the performance of the proposed correction methods: the uniform distribution is non-normal and symmetric, and its MGF has a simple expression; the modified Beta distribution is moderately skewed and  its MGF is quite complicated; the scaled $t$ distribution is symmetric and heavy-tailed, and the MGF does not exist.

 The observed data set available to the naive EB, CEB and BCEB methods is $ \mathcal{O}^* = \{(T_i, \mathbf{Z}_{i1}^*, Y_i): i = 1, \ldots, n\} $.  In Cases III and IV,   we impose the normal error assumption when implementing the CEB method, and thus it is expected to be biased in these situations. The observed data set for the CEB-HL and CEB-HW methods is $ \mathcal{O}^{**} = \{(T_i, \mathbf{Z}_{i1}^*,\mathbf{Z}_{i2}^*, Y_i): i = 1, \ldots, n\}$. Table \ref{Table_ATTEstimation_NormalUniformBetaT}  summarizes the simulation outputs, where we report the bias, standard deviation (SD), and mean square error (MSE) for the naive EB, CEB, BCEB, CEB-HL, and CEB-HW methods.

Table \ref{Table_ATTEstimation_NormalUniformBetaT} shows that naive EB is vulnerable  to covariate measurement error, leading to dramatic  bias and substantially large MSE values across nearly all scenarios. The performance of naive EB is particularly inferior when measurement error becomes large. In comparison,  the proposed correction methods are effective in reducing bias even under challenging scenarios where the errors are moderately skewed or heavy-tailed, or where the error distribution is misspecified.

Among the correction methods, CEB-HW demonstrates superior performance with the lowest bias in all tested scenarios, thereby affirming its robustness against the distributional assumption of measurement errors. In addition, the variance and MSE of CEB-HW is smaller than other methods in almost all scenarios.  The CEB-HL method has comparable performance in Cases I and II where the error distribution is symmetric. In contrast, it  struggles with  moderately skewed and heavy-tailed error distributions in Cases III and IV, suggesting  its application limitation. However, we make a cautionary note that it can be preferred over CEB-HW when some subjects does not have replicated measurements of the error-prone covariates.

Because  CEB and BCEB utilize only one of the two replicated measurements, it is not surprising that they have larger estimation bias, variance, and MSE values than CEB-HW. Although the error distribution is misspecified to be normal when implementing CEB and BCEB in Cases III and  IV, they still reduce substantial bias compared to the naive EB method. In Case II where the uniform error distribution is correctly specified, the CEB method is evidently better than BCEB. In Case I, the  bias and variance of BCEB are  smaller  than those of CEB, which can be explained by that BCEB is numerically more stable than CEB.  Therefore, despite the fact that  BCEB lacks asymptotic guarantees and covariate balancing properties and that it is derived under the normal error assumption, it is a  competitive corrected covariate balancing method.

In conclusion, when replicated measurements are available for most subjects in the study, we recommend CEB-HW; otherwise, we recommend implementing both CEB-HW and CEB-HL and make comparison of their performance. When there is no such additional information and the normal error assumption is reliable, we recommend CEB if covariate balancing and asymptotic  properties are the major concerns, and  recommend BCEB if  numerical stability  is of great importance. If the  error assumption is non-normal and the MGF is simple, we recommend CEB. If the MGF is complicated or does not exist, we recommend BCEB  and  CEB using the misspecified normal MGF.

\begin{table}[H]	\caption{Simulation results for ATT estimation using the naive EB (EB), CEB, BCEB, CEB-HL, and CEB-HW methods.}
	\centering
	\resizebox{0.9\linewidth}{!}{
		\begin{threeparttable}
			\begin{tabular}{ccccccccccccccc}
				\toprule
				\multirow{3}[4]{*}{Error distribution} &       & \multirow{3}[4]{*}{$ \text{Var}(\epsilon) $} &       & \multirow{3}[4]{*}{Metric} &       & \multicolumn{9}{c}{\multirow{2}[2]{*}{Method}} \\
				&       &       &       &       &       & \multicolumn{9}{c}{} \\
				\cmidrule{7-15}          &       &       &       &       &       & EB    &       & CEB  &       & BCEB    &       & CEB-HL   &       & CEB-HW \\
				\midrule
				\multirow{6}[4]{*}{\tabincell{c}{Case I\\~\\ $ N(0,\sigma_{1}^{2}) $}} &       & \multirow{3}[2]{*}{0.10 } &       & Bias  &       & -1.816  &       & 0.038  &       & 0.019  &       & 0.013  &       & 0.013  \\
				&       &       &       & SD    &       & 0.560  &       & 0.657  &       & 0.650  &       & 0.463  &       & 0.459  \\
				&       &       &       & MSE   &       & 3.612  &       & 0.433  &       & 0.422  &       & 0.214  &       & 0.211  \\
				\cmidrule{3-15}          &       & \multirow{3}[2]{*}{0.50 } &       & Bias  &       & -6.104  &       & 0.493  &       & -0.026  &       & 0.335  &       & 0.119  \\
				&       &       &       & SD    &       & 0.904  &       & 1.976  &       & 1.640  &       & 2.085  &       & 1.288  \\
				&       &       &       & MSE   &       & 38.071  &       & 4.145  &       & 2.688  &       & 4.456  &       & 1.670  \\
				\midrule
				\multirow{6}[4]{*}{\tabincell{c}{Case II\\~\\ Uniform}} &       & \multirow{3}[2]{*}{0.10 } &       & Bias  &       & -1.767  &       & 0.043  &       & 0.078  &       & 0.028  &       & 0.025  \\
				&       &       &       & SD    &       & 0.529  &       & 0.608  &       & 0.609  &       & 0.425  &       & 0.450  \\
				&       &       &       & MSE   &       & 3.402  &       & 0.371  &       & 0.377  &       & 0.181  &       & 0.203  \\
				\cmidrule{3-15}          &       & \multirow{3}[2]{*}{0.50 } &       & Bias  &       & -5.916  &       & 0.119  &       & 0.786  &       & 0.070  &       & 0.114  \\
				&       &       &       & SD    &       & 0.886  &       & 1.360  &       & 1.513  &       & 0.955  &       & 1.246  \\
				&       &       &       & MSE   &       & 35.783  &       & 1.862  &       & 2.906  &       & 0.916  &       & 1.564  \\
				\midrule
				\multirow{6}[4]{*}{\tabincell{c}{Case III\\~\\ Modified Beta}} &       & \multirow{3}[2]{*}{0.10 } &       & Bias  &       & -2.108  &       & -0.369  &       & -0.383  &       & -0.373  &       & 0.032  \\
				&       &       &       & SD    &       & 0.576  &       & 0.681  &       & 0.674  &       & 0.513  &       & 0.483  \\
				&       &       &       & MSE   &       & 4.774  &       & 0.599  &       & 0.601  &       & 0.402  &       & 0.234  \\
				\cmidrule{3-15}          &       & \multirow{3}[2]{*}{0.50 } &       & Bias  &       & -7.010  &       & -3.269  &       & -3.163  &       & -3.375  &       & 0.216  \\
				&       &       &       & SD    &       & 0.985  &       & 1.577  &       & 1.519  &       & 1.180  &       & 1.386  \\
				&       &       &       & MSE   &       & 50.111  &       & 13.171  &       & 12.311  &       & 12.783  &       & 1.967  \\
				\midrule
				\multirow{6}[4]{*}{\tabincell{c}{Case IV\\~\\ Scaled $t$}} &       & \multirow{3}[2]{*}{0.10 } &       & Bias  &       & -2.300  &       & -0.676  &       & -0.683  &       & -0.958  &       & 0.108  \\
				&       &       &       & SD    &       & 1.986  &       & 2.520  &       & 2.508  &       & 10.108  &       & 3.314  \\
				&       &       &       & MSE   &       & 9.233  &       & 6.801  &       & 6.748  &       & 102.918  &       & 10.981  \\
				\cmidrule{3-15}          &       & \multirow{3}[2]{*}{0.50 } &       & Bias  &       & -7.204  &       & -4.163  &       & -3.970  &       & -2.774  &       & 0.352  \\
				&       &       &       & SD    &       & 2.420  &       & 5.098  &       & 5.126  &       & 18.831  &       & 7.722  \\
				&       &       &       & MSE   &       & 57.739  &       & 43.296  &       & 42.006  &       & 360.189  &       & 59.639  \\
				\bottomrule
			\end{tabular}%
		\end{threeparttable}
	}
	\label{Table_ATTEstimation_NormalUniformBetaT}%
\end{table} %

\subsection{ASMD comparison \label{sec5.2.2}}

\begin{figure}[H]
	\centering
	\includegraphics[scale=0.55]{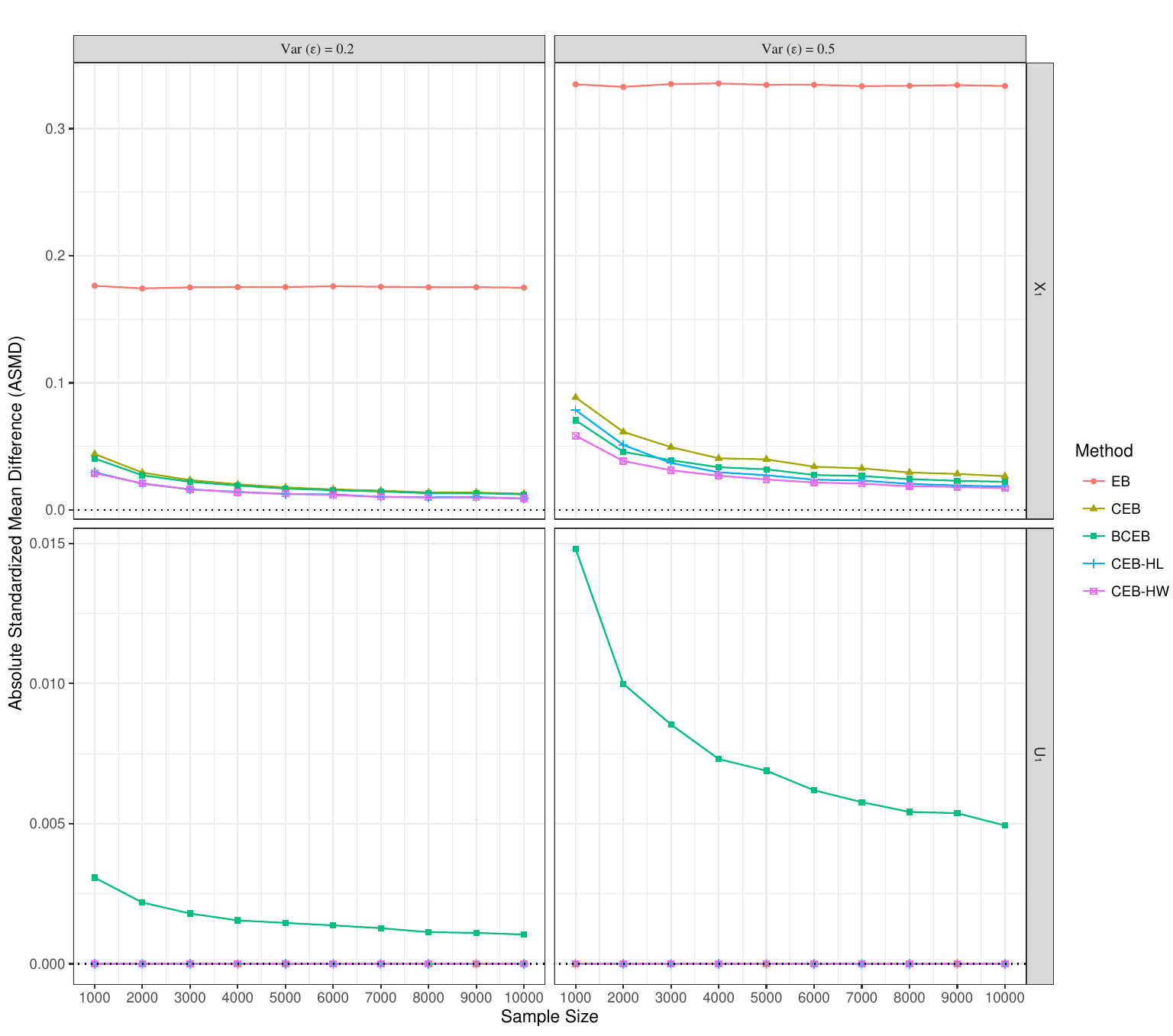}
	\caption{ASMD comparison}
	\label{fig:ASMD}
\end{figure}

We further investigate   the performance of the naive and corrected EB  methods in balancing covariates. The simulation setup is the same as that in Section \ref{sec5.2} with normal errors, except that the sample size $n$ varies from 1000 to 10000  in this subsection, and that  the error variance is  0.2 or 0.5. We report the ASMD values of the unobserved true covariate $X_1$ and the observed true covariate $U_1$.  Figure \ref{fig:ASMD} displays the results.

All correction methods outperform the naive EB in balancing $X_1$. The ASMD values of the correction methods  decrease towards  zero as the sample size increases, suggesting that  the $X_1$ is asymptotically balanced by these methods. The naive EB, CEB, CEB-HL, and CEB-HW methods exactly balance $U_1$, whereas this covariate remains slightly imbalanced by BCEB.

\subsection{Real data analysis \label{sec5.3}}

In this section, we analyze two real data sets: the National Health and Nutrition Examination Survey Data I Epidemiological Follow-up Study (NHEFS) \citep{Robins2020WhatIf} and the AIDS Clinical Trials Group 175 Study (ACTG175) \citep{HIV1996}.

\subsubsection{Analysis of NHEFS data \label{RDA1}}

The NHEFS study  was initiated  in the 1970s.  Baseline demographic and health information  was collected  at the first visit (year 1971-1975) and the follow-up visit (year 1982) for  cigarette smokers. The mortality information was collected in  year 1992. We use a  subset of the NHEFS data comprising  $n= 1138$ subjects. We define ``light smoking'' if  no more than ten cigarettes was consumed per day, whereas  ``heavy smoking'' if cigarette consumption exceeds this threshold. Individuals with successive light smoking behavior at the initial and subsequent visits  are assigned to the treatment group ($T = 1$), and those with successive heavy smoking behavior at both visits are assigned to  the control group ($T = 0$). The outcome variable is defined to be the death indicator, where a value of 1 indicates death by year 1992 and a value of 0 otherwise. The covariates under analysis include age, gender, exercise level, physical activity level, and systolic blood pressure (SBP), all recorded during the first visit. We normalize SBP  to be $\log(SBP - 50)$. Readings of SBP are known to be subject to measurement error \citep{Carroll}.   Measurement error associated with $\log(SBP - 50)$ is assumed to follow the   normal distribution, and its variance was estimated  to be 0.0126  by external study \citep{shu_inverseprobabilitytreatment_2019}.

We are interested in evaluating the effect of  smoking behavior on mortality. We obtain ATT estimates using naive EB, CEB, and BCEB and use
 bootstrap  to compute standard errors (SE) and  the corresponding 95\% confidence intervals (CI). We also report the unadjusted ATT estimate, which is the difference of  average outcomes between  the two  groups. Furthermore, we perform a sensitivity analysis to assess the impact of the magnitude of measurement error variance on the analysis results. By varying the  error variance within the set $\{0.0126, 0.0226, 0.0372, 0.0420\}$, the noise-to-signal ratio (NSR) ranges from 30\% to 100\%.  Table \ref{tab:Result_ATT_boot500_NHEFS} summarizes the results.


\begin{table}[h]
	\centering
	\caption{ATT inference results  for the NHEFS data set.  ``NSR'' refers to the noise-to-signal ratio. ``Estimate'' refers to the ATT point estimate $\times 100$, and ``SE'' and ``95\% CI'' are the bootstrap standard error $\times 100$ and 95 percent bootstrap confidence interval  $\times 100$. }
	\label{tab:Result_ATT_boot500_NHEFS}%
	\resizebox{0.7\textwidth}{!}{
		\begin{threeparttable}
			\begin{tabular}{ccccccccccc}
				\toprule
				Error Variance &       & NSR    &       & Method &       & Estimate &       & SE    &       & 95\% CI \\
				\midrule
				  &       &   &       & Unadjusted &       & 3.738 &       & 2.552 &       & (-1.264, 8.739) \\
				&       &       &       &  EB &       & -1.771 &       & 2.537 &       & (-6.742, 3.201) \\
				\midrule
				\multirow{2}[2]{*}{0.0126} &       & \multirow{2}[2]{*}{30\%} &       & CEB   &       & -2.092 &       & 2.596 &       & (-7.181, 2.997) \\
				&       &       &       & BCEB  &       & -2.097 &       & 2.598 &       & (-7.189, 2.994) \\
				\midrule
				\multirow{2}[2]{*}{0.0226} &       & \multirow{2}[2]{*}{54\%} &       & CEB   &       & -2.597 &       & 2.724 &       & (-7.936, 2.743) \\
				&       &       &       & BCEB  &       & -2.636 &       & 2.740 &       & (-8.006, 2.735) \\
				\midrule
				\multirow{2}[2]{*}{0.0372} &       & \multirow{2}[2]{*}{89\%} &       & CEB   &       & -5.198 &       & 16.042 &       & (-36.641, 26.244) \\
				&       &       &       & BCEB  &       & -6.232 &       & 6.696 &       & (-19.356, 6.892) \\
				\midrule
				\multirow{2}[2]{*}{0.0420} &       & \multirow{2}[2]{*}{100\%} &       & CEB   &       & -9.171 &       & 31.160 &       & (-70.243, 51.901) \\
				&       &       &       & BCEB  &       & -14.940 &       & 19.963 &       & (-54.067, 24.186) \\
				\bottomrule
			\end{tabular}%
		\end{threeparttable}
	}
\end{table}

In contrast to the unadjusted method, the   naive EB, CEB, and BCEB methods produce negative estimates. If light smokers had changed their smoking
behaviour  in the way that they had consumed  more than  ten cigarettes per day, then the  mortality rate by year 1992 will have increased  1.77\% (naive EB) or 2.10\% (CEB and BCEB).  The inference result of CEB or BCEB is insignificant even  when the NSR increases from 30\% to 100\%, although the absolute value of point estimate grows from 2.10\% to as large as 14.95\%. Therefore, the  conclusion of statistical analysis by the correction methods is  consistent  against various magnitudes of measurement error. This phenomenon can be explained by  that the true parameter value for $\log(SBP - 50)$  is  close to zero in this data analysis as suggested by point estimates of naive EB, CEB, and BCEB,  and thus the naive EB method is approximately consistent according to  Theorem \ref{thm2}.

\subsubsection{Analysis of ACTG175 data  \label{RDA2}}

We employ the unadjusted, naive EB, and the proposed corrected EB methods to the ACTG175 study. We analyze a subset of $ n= 2094 $ participants with two baseline measurements of CD4 counts prior to assignment of  nucleoside  monotherapy or combination therapy. We are interested in evaluating whether participants who had benefited from   their therapies in the short term will experience   a prolonged  beneficial effect.

The CD4 count plays a critical role in assessing HIV patient immunity. It was measured twice at the first visit just prior to  therapy assignments to the participants, and was measured again $20 \pm 5$  weeks later at the second visit. Almost all participants survived to the second visit.   We define the treatment indicator to capture the variation in
   the CD4 counts at the first visit (cd40) and the subsequent visit (cd420): $T=I(cd420 - cd40 > 0)$. If $T=1$ for an   individual, then it indicates an increment of the CD4 cell count  20 weeks after his/her therapy,  thereby experiencing a short-term beneficial effect. We define the outcome variable to be the occurrence of an AIDS-associated event several months after the second visit. We consider the number of months  $D$   to be $6, 12$, or 30.

The error-free observed covariate vector $\bU$ consists of age, weight, Karnofsky score, CD8 count, gender, race,  antiretroviral history, among others. The CD4 count  is susceptible  to measurement error due to daily and biological fluctuations \citep{Tsiatis1995}.  Let $X$ be the true  log-transformed baseline CD4 count. Let  $X_{1}^*$ and $X_{2}^*$ be the replicated measurements obtained  in chronological order. The CEB-HL and CEB-HW methods utilize both  replicated measurements for ATT estimation, whereas the  naive EB, CEB, and BCEB   set $X^*=X_{2}^*$.
The CEB method is implemented assuming  normally distributed measurement error \citep{Tsiatis1995}, where the error variance is estimated by replicated measurements. Table \ref{Table_ATT100SD100CI100_ACTG175} presents the results, where the ATT estimate, SE and 95\% CI are reported for all methods.

All methods show that the short-term CD4 count elevation   has a long-term beneficial effect on reducing the chance of long-term AIDS-related event occurrence in the treated group. The naive EB-based ATT estimate is smaller than the unadjusted ATT estimate by $0.60\sigma$ to $0.70\sigma$, where $\sigma$ is the bootstrap standard error of the unadjusted ATT estimate. In comparison, the ATT estimates by the corrected methods are uniformly smaller. For example, the  CEB-HW-based ATT estimate is smaller than the unadjusted ATT estimate by $1.38\sigma$ to $1.61\sigma$. Therefore, the naive EB method evidently underestimates the magnitude of ATT.

\begin{table}[h] 	
	\centering \caption{ATT inference results  for the ACTG175 data set. ``Estimate'' refers to the ATT point estimate $\times 100$, and ``SE'' and ``95\% CI'' are the bootstrap standard error $\times 100$ and 95 percent confidence interval estimate $\times 100$.}
	\resizebox{1.00\textwidth}{!}{
		\begin{threeparttable}
			\begin{tabular}{ccccccccccccccc}
				\toprule
				Month &       &       &       & Unadjusted &       & EB &       & CEB   &       & BCEB  &       & CEB-HL    &       & CEB-HW \\
				\midrule
				\multirow{3}[2]{*}{6 } &       & Estimate &       & -7.874 &       & -8.597 &       & -9.294 &       & -9.316 &       & -9.117 &       & -9.291 \\
				&       & SE    &       & 1.027 &       & 1.189 &       & 1.314 &       & 1.316 &       & 1.261 &       & 1.280 \\
				&       & 95\% CI &       & (-9.888, -5.860) &       & (-10.929, -6.266) &       & (-11.869, -6.720) &       & (-11.896, -6.737) &       & (-11.589, -6.645) &       & (-11.800, -6.783) \\
				\midrule
				\multirow{3}[2]{*}{12} &       & Estimate &       & -12.430 &       & -13.371 &       & -14.251 &       & -14.276 &       & -14.322 &       & -14.570 \\
				&       & SE    &       & 1.333 &       & 1.491 &       & 1.628 &       & 1.629 &       & 1.575 &       & 1.588 \\
				&       & 95\% CI &       & (-15.043, -9.817) &       & (-16.293, -10.449) &       & (-17.442, -11.060) &       & (-17.469, -11.083) &       & (-17.409, -11.236) &       & (-17.683, -11.457) \\
				\midrule
				\multirow{3}[2]{*}{30 } &       & Estimate &       & -17.611 &       & -18.755 &       & -20.017 &       & -20.053 &       & -19.840 &       & -20.171 \\
				&       & SE    &       & 1.831 &       & 1.885 &       & 2.014 &       & 2.013 &       & 1.917 &       & 1.919 \\
				&       & 95\% CI &       & (-21.200, -14.021) &       & (-22.450, -15.061) &       & (-23.964, -16.071) &       & (-23.998, -16.107) &       & (-23.596, -16.083) &       & (-23.931, -16.410) \\
				\bottomrule
			\end{tabular}%
		\end{threeparttable}
	}
	\label{Table_ATT100SD100CI100_ACTG175}%
\end{table}

\section{Conclusion \label{sec:conclution}}

Measurement error  poses  great challenges in empirical research, potentially biasing treatment effect estimates and invalidating inference conclusions. Previous studies in  \cite{McCaffrey} and \cite{Lockwood} demonstrated that neglecting measurement error in PSW and matching methods  introduces substantial bias. In this article, we conduct a systematic bias analysis of the influence of measurement error on covariate balancing and treatment effect estimation. We derive various bias formulas and  bounds,  elucidating the limitation of the  EB and CBPS frameworks. To address the challenges, we propose  a class of correction strategies, most of which maintain  finite-sample balance of the observable true covariates and achieve large-sample balance of the unobservable true covariates. The proposed measurement error  correction methods are   functional  in the sense that we do not make distributional assumptions on the covariates \citep{Carroll,Yan,Yan3}.
We derive  the asymptotic properties  of the CEB method.
The simulation studies reveal that  measurement error has a dramatic impact on covariate balance and confirm that  the proposed methods successfully adjust for measurement error.  The impact of misspecification of the measurement error model  can be further explored \citep{YiYan2021}.

Although we primarily focus on estimating the average treatment effect on the treated, our correction strategies  can be easily adapted to other causal estimands, such as the average treatment effect or the average treatment effect on the control. For studies involving  categorical or survival outcomes, alternative estimands such as relative risk, odds ratio, or difference  in  the restricted mean survival times can be considered.

Future research efforts are needed to balance mismeasured covariates for complex data structure such as  high-dimensional and longitudinal data collected from various study designs \citep[e.g.,][]{YanZhouCai2017,LiangYan2022,YiYanLiaoSpiegelman}, for mediation analysis  \citep{YanRenLeon2023}, and for statistical learning of  individualized treatment rules \citep[e.g.,][]{LiZhouWuYan2023}.

\section*{Supplementary Material}
The  Supplementary Material includes the proofs and regularity assumptions.

\section*{Acknowledgement}
Ying Yan's research was partially supported by the National Natural Science Foundation of China (Grant No. 12292984, 11901599).

\bibliographystyle{ecta}
\small{
\bibliography{paper-ref}

\begin{thebibliography}{44}
\newcommand{\enquote}[1]{``#1''}
\expandafter\ifx\csname natexlab\endcsname\relax\def\natexlab#1{#1}\fi

\bibitem[\protect\citeauthoryear{Carroll, Ruppert, Stefanski, and Crainiceanu}{Carroll et~al.}{2006}]{Carroll}
\textsc{Carroll, R.~J., D.~Ruppert, L.~A. Stefanski, and C.~M. Crainiceanu} (2006): \emph{Measurement Error in Nonlinear Models: A Modern Perspective}, Chapman and Hall/CRC, Boca Raton, second ed.

\bibitem[\protect\citeauthoryear{Chan, Yam, and Zhang}{Chan et~al.}{2016}]{chan2016globally}
\textsc{Chan, K. C.~G., S.~C.~P. Yam, and Z.~Zhang} (2016): \enquote{Globally efficient non-parametric inference of average treatment effects by empirical balancing calibration weighting,} \emph{Journal of the Royal Statistical Society: Series B (Statistical Methodology)}, 78, 673--700.

\bibitem[\protect\citeauthoryear{Dai and Yan}{Dai and Yan}{2024}]{DaiYan2022}
\textsc{Dai, Y. and Y.~Yan} (2024): \enquote{Mahalanobis balancing: a multivariate perspective on approximate covariate balancing,} \emph{Scandinavian Journal of Statistics}, Published Online.

\bibitem[\protect\citeauthoryear{Fan, Imai, Lee, Liu, Ning, and Yang}{Fan et~al.}{2023}]{fan2016improving}
\textsc{Fan, J., K.~Imai, I.~Lee, H.~Liu, Y.~Ning, and X.~Yang} (2023): \enquote{Optimal covariate balancing conditions in propensity score estimation,} \emph{Journal of Busines \& Economic Statistics}, 41, 97--110.

\bibitem[\protect\citeauthoryear{Fong, Hazlett, and Imai}{Fong et~al.}{2018}]{FongHazlettImai2018}
\textsc{Fong, C., C.~Hazlett, and K.~Imai} (2018): \enquote{Covariate balancing propensity score for a continuous treatment: Application to the efficacy of political advertisements,} \emph{The Annals of Applied Statistics}, 12, 156--177.

\bibitem[\protect\citeauthoryear{Hainmueller}{Hainmueller}{2012}]{hainmueller2012entropy}
\textsc{Hainmueller, J.} (2012): \enquote{Entropy balancing for causal effects: A multivariate reweighting method to produce balanced samples in observational studies,} \emph{Political Analysis}, 20, 25--46.

\bibitem[\protect\citeauthoryear{Hammer, Katzenstein, Hughes, Gundacker, Schooley, Haubrich, Henry, Lederman, Phair, Niu, Hirsch, and Merigan}{Hammer et~al.}{1996}]{HIV1996}
\textsc{Hammer, S.~M., D.~A. Katzenstein, M.~D. Hughes, H.~Gundacker, R.~T. Schooley, R.~H. Haubrich, W.~K. Henry, M.~M. Lederman, J.~P. Phair, M.~Niu, M.~S. Hirsch, and T.~C. Merigan} (1996): \enquote{A trial comparing nucleoside monotherapy with combination therapy in HIV-infected adults with CD4 cell counts from 200 to 500 per cubic millimeter,} \emph{New England Journal of Medicine}, 335, 1081--1090, pMID: 8813038.

\bibitem[\protect\citeauthoryear{Hern{\'a}n and Robins}{Hern{\'a}n and Robins}{2020}]{Robins2020WhatIf}
\textsc{Hern{\'a}n, M. and J.~Robins} (2020): \emph{Causal Inference: What If}, Chapman and Hall/CRC, Boca Raton.

\bibitem[\protect\citeauthoryear{Hirano, Imbens, and Ridder}{Hirano et~al.}{2003}]{hirano2003efficient}
\textsc{Hirano, K., G.~W. Imbens, and G.~Ridder} (2003): \enquote{Efficient estimation of average treatment effects using the estimated propensity score,} \emph{Econometrica}, 71, 1161--1189.

\bibitem[\protect\citeauthoryear{Hu and Lin}{Hu and Lin}{2004}]{hu2004semiparametric}
\textsc{Hu, C. and D.~Y. Lin} (2004): \enquote{Semiparametric failure time regression with replicates of mismeasured covariates,} \emph{Journal of the American Statistical Association}, 99, 105--118.

\bibitem[\protect\citeauthoryear{Huang and Wang}{Huang and Wang}{2000}]{huang2000cox}
\textsc{Huang, Y. and C.~Wang} (2000): \enquote{Cox regression with accurate covariates unascertainable: A nonparametric-correction approach,} \emph{Journal of the American Statistical Association}, 95, 1209--1219.

\bibitem[\protect\citeauthoryear{Huang and Wang}{Huang and Wang}{2001}]{HuangWang2001}
\textsc{Huang, Y. and C.~Y. Wang} (2001): \enquote{Consistent functional methods for logistic regression with errors in covariates,} \emph{Journal of the American statistical Association}, 96, 1469--1482.

\bibitem[\protect\citeauthoryear{Imai and Ratkovic}{Imai and Ratkovic}{2014}]{imai2014covariate}
\textsc{Imai, K. and M.~Ratkovic} (2014): \enquote{Covariate balancing propensity score,} \emph{Journal of the Royal Statistical Society: Series B (Statistical Methodology)}, 76, 243--263.

\bibitem[\protect\citeauthoryear{Imai and Ratkovic}{Imai and Ratkovic}{2015}]{ImaiRatkovic2015}
---\hspace{-.1pt}---\hspace{-.1pt}--- (2015): \enquote{Robust estimation of inverse probability weights for marginal structural models,} \emph{Journal of the American Statistical Association}, 110, 1013--1023.

\bibitem[\protect\citeauthoryear{Imbens and Rubin}{Imbens and Rubin}{2015}]{imbens2015causal}
\textsc{Imbens, G.~W. and D.~B. Rubin} (2015): \emph{Causal Inference in Statistics, Social, and Biomedical Sciences}, Cambridge University Press.

\bibitem[\protect\citeauthoryear{Josey, Juarez-Colunga, Yang, and Ghosh}{Josey et~al.}{2021}]{Josey2021}
\textsc{Josey, K.~P., E.~Juarez-Colunga, F.~Yang, and D.~Ghosh} (2021): \enquote{A framework for covariate balance using Bregman distances,} \emph{Scandinavian Journal of Statistics}, 48, 790--816.

\bibitem[\protect\citeauthoryear{Kallus and Santacatterina}{Kallus and Santacatterina}{2021}]{Kallus2021}
\textsc{Kallus, N. and M.~Santacatterina} (2021): \enquote{Optimal balancing of time-dependent confounders for marginal structural models,} \emph{Journal of Causal Inference}, 9, 345--369.

\bibitem[\protect\citeauthoryear{Kang and Schafer}{Kang and Schafer}{2007}]{kang2007demystifying}
\textsc{Kang, J.~D. and J.~L. Schafer} (2007): \enquote{Demystifying double robustness: A comparison of alternative strategies for estimating a population mean from incomplete data,} \emph{Statistical Science}, 22, 523--539.

\bibitem[\protect\citeauthoryear{Lee, Yang, Dong, Wang, Zeng, and Cai}{Lee et~al.}{2023}]{dong2020integrative}
\textsc{Lee, D., S.~Yang, L.~Dong, X.~Wang, D.~Zeng, and J.~Cai} (2023): \enquote{Improving trial generalizability using observational studies,} \emph{Biometrics}, 79, 1213--1225.

\bibitem[\protect\citeauthoryear{Li, Zhou, Wu, and Yan}{Li et~al.}{2023}]{LiZhouWuYan2023}
\textsc{Li, X., Q.~Zhou, Y.~Wu, and Y.~Yan} (2023): \enquote{Multicategory matched learning for estimating optimal individualized treatment rules in observational studies with application to a hepatocellular carcinoma study,} \emph{Arxiv preprint}, arXiv:2302.05287.

\bibitem[\protect\citeauthoryear{Liang and Yan}{Liang and Yan}{2022}]{LiangYan2022}
\textsc{Liang, W. and Y.~Yan} (2022): \enquote{Empirical likelihood-based estimation and inference in randomized controlled trials with high-dimensional covariates,} \emph{Statistics and Its Interface}, 15, 283--301.

\bibitem[\protect\citeauthoryear{Lockwood and McCaffrey}{Lockwood and McCaffrey}{2016}]{Lockwood}
\textsc{Lockwood, J.~R. and D.~F. McCaffrey} (2016): \enquote{Matching and weighting with functions of error-prone covariates for causal inference,} \emph{Journal of the American Statistical Association}, 111, 1831--1839.

\bibitem[\protect\citeauthoryear{Maathuis, Drton, Lauritzen, and Wainwright}{Maathuis et~al.}{2018}]{maathuis2018handbook}
\textsc{Maathuis, M., M.~Drton, S.~Lauritzen, and M.~Wainwright} (2018): \emph{Handbook of Graphical Models}, CRC Press.

\bibitem[\protect\citeauthoryear{McCaffrey, Lockwood, and Setodji}{McCaffrey et~al.}{2013}]{McCaffrey}
\textsc{McCaffrey, D.~F., J.~R. Lockwood, and C.~M. Setodji} (2013): \enquote{Inverse probability weighting with error-prone covariates,} \emph{Biometrika}, 100, 671--680.

\bibitem[\protect\citeauthoryear{Ogburn and Vanderweele}{Ogburn and Vanderweele}{2013}]{ogburn2013bias}
\textsc{Ogburn, E.~L. and T.~J. Vanderweele} (2013): \enquote{Bias attenuation results for nondifferentially mismeasured ordinal and coarsened confounders,} \emph{Biometrika}, 100, 241--248.

\bibitem[\protect\citeauthoryear{Rosenbaum and Rubin}{Rosenbaum and Rubin}{1983}]{rosenbaum1983central}
\textsc{Rosenbaum, P.~R. and D.~B. Rubin} (1983): \enquote{The central role of the propensity score in observational studies for causal effects,} \emph{Biometrika}, 70, 41--55.

\bibitem[\protect\citeauthoryear{Shu and Yi}{Shu and Yi}{2019}]{shu_inverseprobabilitytreatment_2019}
\textsc{Shu, D. and G.~Y. Yi} (2019): \enquote{Inverse probability of treatment weighted estimation of causal parameters in the presence of error-contaminated and time-dependent confounders,} \emph{Biometrical Journal}, 61, 1507--1525.

\bibitem[\protect\citeauthoryear{Stefanski and Carroll}{Stefanski and Carroll}{1987}]{StefanskiCarroll1987}
\textsc{Stefanski, L.~A. and R.~J. Carroll} (1987): \enquote{Conditional scores and optimal scores in generalized linear measurement error models,} \emph{Biometrika}, 74, 703--716.

\bibitem[\protect\citeauthoryear{Tan}{Tan}{2020}]{tan2020regularized}
\textsc{Tan, Z.} (2020): \enquote{Regularized calibrated estimation of propensity scores with model misspecification and high-dimensional data,} \emph{Biometrika}, 107, 137--158.

\bibitem[\protect\citeauthoryear{Tsiatis, Degruttola, and Wulfsohn}{Tsiatis et~al.}{1995}]{Tsiatis1995}
\textsc{Tsiatis, A.~A., V.~Degruttola, and M.~S. Wulfsohn} (1995): \enquote{Modeling the relationship of survival to longitudinal data measured with error, applications to survival and CD4 counts in patients with AIDS,} \emph{Journal of the American Statistical Association}, 90, 27--37.

\bibitem[\protect\citeauthoryear{Wang and Zubizarreta}{Wang and Zubizarreta}{2020}]{wang2020minimal}
\textsc{Wang, Y. and J.~R. Zubizarreta} (2020): \enquote{Minimal dispersion approximately balancing weights: asymptotic properties and practical considerations,} \emph{Biometrika}, 107, 93--105.

\bibitem[\protect\citeauthoryear{Yan and Ren}{Yan and Ren}{2023}]{YanRen}
\textsc{Yan, Y. and M.~Ren} (2023): \enquote{Consistent inverse probability of treatment weighted estimation of the average treatment effect with mismeasured time-dependent confounders,} \emph{Statistics in Medicine}, 42, 517--535.

\bibitem[\protect\citeauthoryear{Yan, Ren, and de~Leon}{Yan et~al.}{2023}]{YanRenLeon2023}
\textsc{Yan, Y., M.~Ren, and A.~de~Leon} (2023): \enquote{Measurement error correction in mediation analysis under the additive hazards model,} \emph{Communications in Statistics-Simulation and Computation}, Published Online.

\bibitem[\protect\citeauthoryear{Yan and Yi}{Yan and Yi}{2015}]{Yan}
\textsc{Yan, Y. and G.~Y. Yi} (2015): \enquote{A corrected profile likelihood method for survival data with covariate measurement error under the Cox model,} \emph{Canadian Journal of Statistics}, 43, 454--480.

\bibitem[\protect\citeauthoryear{Yan and Yi}{Yan and Yi}{2016{\natexlab{a}}}]{Yan3}
---\hspace{-.1pt}---\hspace{-.1pt}--- (2016{\natexlab{a}}): \enquote{Analysis of error-prone survival data under additive hazards models: Measurement error effects and adjustments,} \emph{Lifetime Data Analysis}, 22, 321--342.

\bibitem[\protect\citeauthoryear{Yan and Yi}{Yan and Yi}{2016{\natexlab{b}}}]{Yan2}
---\hspace{-.1pt}---\hspace{-.1pt}--- (2016{\natexlab{b}}): \enquote{A class of functional methods for error-contaminated survival data under additive hazards models with replicate measurements,} \emph{Journal of the American Statistical Association}, 111, 684--695.

\bibitem[\protect\citeauthoryear{Yan, Zhou, and Cai}{Yan et~al.}{2017}]{YanZhouCai2017}
\textsc{Yan, Y., H.~Zhou, and J.~Cai} (2017): \enquote{Improving efficiency of parameter estimation in case-cohort studies with multivariate failure time data,} \emph{Biometrics}, 73, 1042--1052.

\bibitem[\protect\citeauthoryear{Yi}{Yi}{2017}]{Yibook}
\textsc{Yi, G.~Y.} (2017): \emph{Statistical Analysis with Measurement Error or Misclassification}, New York, NY: Springer.

\bibitem[\protect\citeauthoryear{Yi and Yan}{Yi and Yan}{2021}]{YiYan2021}
\textsc{Yi, G.~Y. and Y.~Yan} (2021): \enquote{Estimation and hypothesis testing with error-contaminated survival data under possibly misspecified measurement error models,} \emph{Canadian Journal of Statistics}, 49, 853--874.

\bibitem[\protect\citeauthoryear{Yi, Yan, Liao, and Spiegelman}{Yi et~al.}{2019}]{YiYanLiaoSpiegelman}
\textsc{Yi, G.~Y., Y.~Yan, X.~Liao, and D.~Spiegelman} (2019): \enquote{Parametric regression analysis with covariate misclassification in main study/validation study designs,} \emph{International Journal of Biostatistics,}, 15, 20170002.

\bibitem[\protect\citeauthoryear{Zhao}{Zhao}{2019}]{Zhao2019}
\textsc{Zhao, Q.} (2019): \enquote{Covariate balancing propensity score by tailored loss functions,} \emph{The Annals of Statistics}, 47, 965--993.

\bibitem[\protect\citeauthoryear{Zhao and Percival}{Zhao and Percival}{2017}]{zhao2016entropy}
\textsc{Zhao, Q. and D.~Percival} (2017): \enquote{Entropy balancing is doubly robust,} \emph{Journal of Causal Inference}, 5, 20160010.

\bibitem[\protect\citeauthoryear{Zhou and Wodtke}{Zhou and Wodtke}{2020}]{ZhouWodtke2020}
\textsc{Zhou, X. and G.~T. Wodtke} (2020): \enquote{Residual balancing: A method of constructing weights for marginal structural models,} \emph{Political Analysis}, 28, 487--506.

\bibitem[\protect\citeauthoryear{Zubizarreta}{Zubizarreta}{2015}]{zubizarreta2015stable}
\textsc{Zubizarreta, J.~R.} (2015): \enquote{Stable weights that balance covariates for estimation with incomplete outcome data,} \emph{Journal of the American Statistical Association}, 110, 910--922.

\end{thebibliography}
}

\end{document}